\documentclass[twocolumn]{aastex63}
\usepackage{enumitem}
\usepackage{amsmath}
\usepackage{eqnarray}
\usepackage{tablefootnote}
\usepackage{pifont}

\newcommand{\Ms}{M$_{\odot}$}

\def\gsim{\mathrel{\raise.5ex\hbox{$>$}\mkern-14mu
             \lower0.6ex\hbox{$\sim$}}}

\def\lsim{\mathrel{\raise.3ex\hbox{$<$}\mkern-14mu
             \lower0.6ex\hbox{$\sim$}}}

\begin{document}

\title{Dust formation in AGN winds}
\shorttitle{Dust formation in AGNs}
\author{Arkaprabha Sarangi}
\affiliation{Observational Cosmology Lab, NASA Goddard Space Flight Center, Mail Code 665, Greenbelt, MD 20771, USA}
\affiliation{CRESST II/CUA/GSFC}
\email{arkaprabha.sarangi@nasa.gov}

\author{Eli Dwek}
\affiliation{Observational Cosmology Lab, NASA Goddard Space Flight Center, Mail Code 665, Greenbelt, MD 20771, USA}

\author{Demos Kazanas}
\affiliation{Gravitational Astrophysics Lab, NASA Goddard Space Flight Center, Mail Code 663, Greenbelt, MD 20771, USA}

\begin{abstract}
Infrared observations of active galactic nucleus (AGN) reveal emission from the putative dusty circumnuclear ``torus" invoked by AGN unification, that is heated up by radiation from the central accreting black hole (BH). The strong 9.7 and 18 $\mu$m silicate features observed in the AGN spectra both in emission and absorption, further indicate the presence of such dusty environments. We present detailed calculations of the chemistry of silicate dust formation in AGN accretion disk winds. The winds considered herein are magnetohydrodynamic (MHD) winds driven off the entire accretion disk domain that extends from the BH vicinity to the radius of BH influence, of order $\sim 1-100$ pc depending on the mass of the resident BH. Our results indicate that these winds provide conditions conducive to the formation of significant amounts of dust, especially for objects accreting close to their Eddington limit, making AGN a significant source of dust in the universe, especially for luminous quasars. Our models justify the importance of a $r^{-1}$ density law in the winds for efficient formation and survival of dust grains. The dust production rate scales linearly with the mass of the central BH and varies as a power law of index between 2 to 2.5 with the dimensionless mass accretion rate. The resultant distribution of the dense dusty gas resembles a toroidal shape, with high column density and optical depths along the equatorial viewing angles, in agreement with the AGN unification picture. 
\end{abstract}


\keywords{ISM: dust, extinction; stars: black holes, winds, outflows; galaxies: quasars: general}
\section{Introduction}

The presence of dust in galaxies has a profound effect on their spectral appearance. Dust manifests itself in the absorption and scattering of starlight, in the production of interstellar polarization, X-ray haloes, and echoes around variable sources, and in the depletion of refractory elements from the ISM gas \citep{li2005, draine2003, zub04}. On a cosmological scale, dust reprocesses a significant fraction of all gravitational and nuclear energy releases throughout cosmic history into IR radiation, generating the cosmic IR background \citep{hauser2001, dwek2013}.

Despite the many manifestations of dust, its nature, origin, and evolution in  galaxies are still poorly understood. For instance, a detailed study of the nearby Magellanic Clouds revealed that the dust formation rate in the ejecta of core collapse supernovae and the quiescent winds of low mass stars is significantly lower than its destruction rate by blast waves from supernova remnants. Therefore, the presence of dust in the LMC suggests that we are unable to account for the total amount of dust observed in the galaxy \citep{temim2015}. In local galaxies, where the primary sites of dust formation are known to be quiescent winds from the asymptotic giant branch (AGB) stars, the disparity of the rates between dust and star formation implies also the necessity of an additional source of dust \citep{dwe11}.

A large mass of dust ($\sim$ 10$^8$ \Ms) is reported to be present in a number of  high redshift ($z > 6$) quasars \citep{ber03}, which implies a rapid rate of dust formation in the short $\simeq 0.7$ Gyr time span between the reionization redshift, $z_r \simeq 17$ and $z \simeq 6$.  Supernovae, resulting from rapidly evolving supermassive stars, are considered to be the primary source of refractory elements and dust in the early universe.  

\cite{elvis2002} proposed that dust may be produced in outflows from the broad line region of an AGN, noting that the conditions (density and temperature) in the AGN Broad Line Region (BLR) clouds are similar to those in the winds of AGB stars, where most of the galactic dust is thought to be made.  From this argument and the thermodynamic conditions of AGN outflows they estimated a dust production rate of $\sim$ 10$^{-2}$ \Ms\ yr$^{-1}$. More recently,  \cite{maiolino2006} invoking the \cite{elvis2002} estimates, suggested that they could account for all dust in the high redshift (z = 6.4) quasar SDSSJ1148+52, while \cite{pipino2006} argued that dust formation would only be important in the earlier state of AGN life. 

Besides their potential role as dust sources at high redshifts, local AGN are known to contain significant quantities of dust, considering that structures, referred to as ``dusty tori", are thought to be ubiquitous features of the AGN morphology, closely related to the unification of Seyfert 1,2 \citep{antonucci1993} and the Broad and Narrow Line Radio Galaxies \citep{barthel1989}. According to AGN unification, these distinct classes represent single entities, distinguished only by their angle of inclinations with respect to the viewer's line of sight. Dust is presumably present in toroidal structures obstructing direct view of the AGN central regions for observers at sufficiently large inclination positions \citep{mason2015}.

The sizes of the ``dusty tori" ($R \sim 1$ pc) are much too small to resolve angularly, even for the closest AGN. Spectroscopically, the presence of dust in the galaxies with an AGN has been well documented over the past couple of decades \citep{almeida2017}. The fact that the AGN IR luminosity is similar to that of UVO - X-ray one, in low inclination objects where the central engine is directly visible, implies that the tori subtend a significant fraction of the continuum source luminosity. For observers at higher inclination, the obscuration of the central regions may bring most of the observed luminosity to the IR.
In this respect, several radiative transfer models in the past couple of decades have focused on the study of infrared emission and absorption from this dusty toroidal gas as a function viewing angle \citep{pier1992, rowanrobinson1995, fritz2006, honig2006, stalevski2012}.

Silicates are known to be the primary constituent of dust in the AGN torus \citep{stenholm1994, srinivasan2017}. In the last decade, Spitzer Space Telescope has detected strong emission features of silicate dust present in quasars \citep{hao2005, siebenmorgen2005, sturm2005}. For the Seyfert 2  NGC~1068, silicate features have been reported in absorption \citep{mason2006} as well as emission \citep{enrique2018, enrique2018b}. Detection of emission features in a type 2 AGN could indicate to a possible clumpy nature of the gas in the outflows \citep{elitzur2007, tristram2007, nenkova2008}.

The dust present in the AGN tori can be pre-existing dust from the ISM, newly formed dust in the winds or a combination of both. The detection of the 9.7 and 18 $\mu$m silicate features, reported in the AGN spectra \citep{shi2006}, are characteristic to the tetrahedral Si-O bonds which naturally form in a high temperature condensation sequences. In as much as the formation of these bonds requires high activation energy, then these emission features suggest the in situ formation of silicates in the wind. Also, the crystallinity of the dust grains in AGNs is  found to be higher compared to the upper limit in the local ISM \citep{sturm2005, srinivasan2017},  arguing in the favor of the formation of new dust grains in the AGN.

In spite of recent developments in the field, the structure and origin of the AGN tori is still uncertain, as is the origin of dust in them. Their most vexing issue is their large vertical extent, of height $H$ comparable to their distance to the AGN center $R$. Under hydrostatic equilibrium this implies random velocities comparable to the Keplerian one ($V_K$) at distance $R \sim 1-10$ pc, $V_K \sim 300-1000$ km/s (for a BH mass $M \simeq 10^8 M_{\odot}$), which is in stark disagreement with their low gas and dust temperatures ($10 - 100$ K) that imply velocities $\langle V^2 \rangle^{1/2} \sim 1$ km/s. According to certain models \citep{hopkins2012} the dusty tori represent gas inflowing from the galaxies' larger radii.  Alternatively, it has been suggested that they represent MHD winds outflowing from the outer regions of the AGN accretion disks \citep{konigl1994, elitzur2006, keating2012}, and therefore are not subject to the constraints of their internal velocities, implied by assuming hydrostatic equilibrium.

The MHD wind-view is strengthened by the discovery that approximately 50\% of AGN exhibit UV and X-ray \emph{blue-shifted} absorption line features \citep{crenshaw2003}, indicating the generic presence of outflows in AGN. Of particular interest is the absorption features of the AGN X-ray spectra, because they span a much wider range of ionization states thereby probing a wider range of outflow conditions. At the same time, because of the high ionization states of their plasma, they preclude winds driven by line radiation pressure akin to those of O-star winds. These facts, therefore, argue in favor of MHD driven winds \citep{blandford1982, conto94}. These winds, due to their self-similar structure, span a large range of radii (and hence possibly plasma ionization states) and do not rely on radiation pressure for their launch. The study of the polarization of dust emission in Cygnus A also supports the magnetic field dominated wind structure \citep{enrique2018b}.

The wide range of ionization states of X-ray absorbers (equivalently the large distance of absorbing plasma from the ionizing source of radiation), provides a means of estimating the plasma density along the observers' line of sight \citep{george1998}, by measuring the absorbing column of each transition. Such analyses by \cite{holczer2007} and \cite{behar2009} showed that the wind density profile is close to $n(r) \propto r^{-1}$ rather than $r^{-2}$, generally expected in a radiation pressure driven wind. Furthermore, it should be noted that these shallow density profiles are consistent with the winds proposed by \cite{conto94} rather than those of \cite{blandford1982} which imply $n(r) \propto r^{-3/2}$. The photoionization structure of these MHD winds were modeled by \cite{fukumura2010,fukumura2018} and was shown to be consistent with the propositions of AGN unification.

Motivated by these developments  we study the formation of silicate dust in AGN MHD accretion disk winds. With a well defined background model for the MHD winds that represents the AGN tori, we formulate a simple, generic dust evolution model to study the chemistry of dust formation in these winds. Our aim is to quantify the following: (a) the pathways of silicate dust formation in the outflows, (b) the two dimensional distribution of dust density and temperatures, (c) the dust grain sizes, (d) the dust column density present along various line of sight, (e) the rate of silicate dust formation and (f) the global dusty structure and dust formation budget of the associated AGN tori.


The paper is arranged in the following order. In Section~\ref{sec_phmodel}, we  discuss the geometry and input constraints on the physical model of the winds, along with the parameters related to the central source. Following that, in Section~\ref{sec_results} we discuss the evolution and distribution of the gas temperature (Section~\ref{sec_gasT}), the dust temperature (Section~\ref{sec_dustT}) and the characteristics of different regions in the winds (Section~\ref{sec_regions}), that are relevant to dust formation. The rate and mechanism of nucleation of molecules and dust precursor clusters are presented in Section~\ref{nucleation}. The estimated density of dust grains and formation of the torus is explained in Section~\ref{sec_dustdensity}. The distribution of grain sizes, and the dust to gas mass ratio is calculated in Section~\ref{sec_grainsizes}, while the total dust mass and the optical depths are discussed in  Section~\ref{sec_dustmass}. In Section \ref{dependancies} we show how the calculated results depend on the variation of the mass accretion rate and the mass of the BH. Finally in Section~\ref{sec_conclusion}, we summarize our findings and discuss the dependencies of the model on various physical parameters.

\begin{figure*}
\vspace*{0.3cm}
\centering
\includegraphics[width=3.5in]{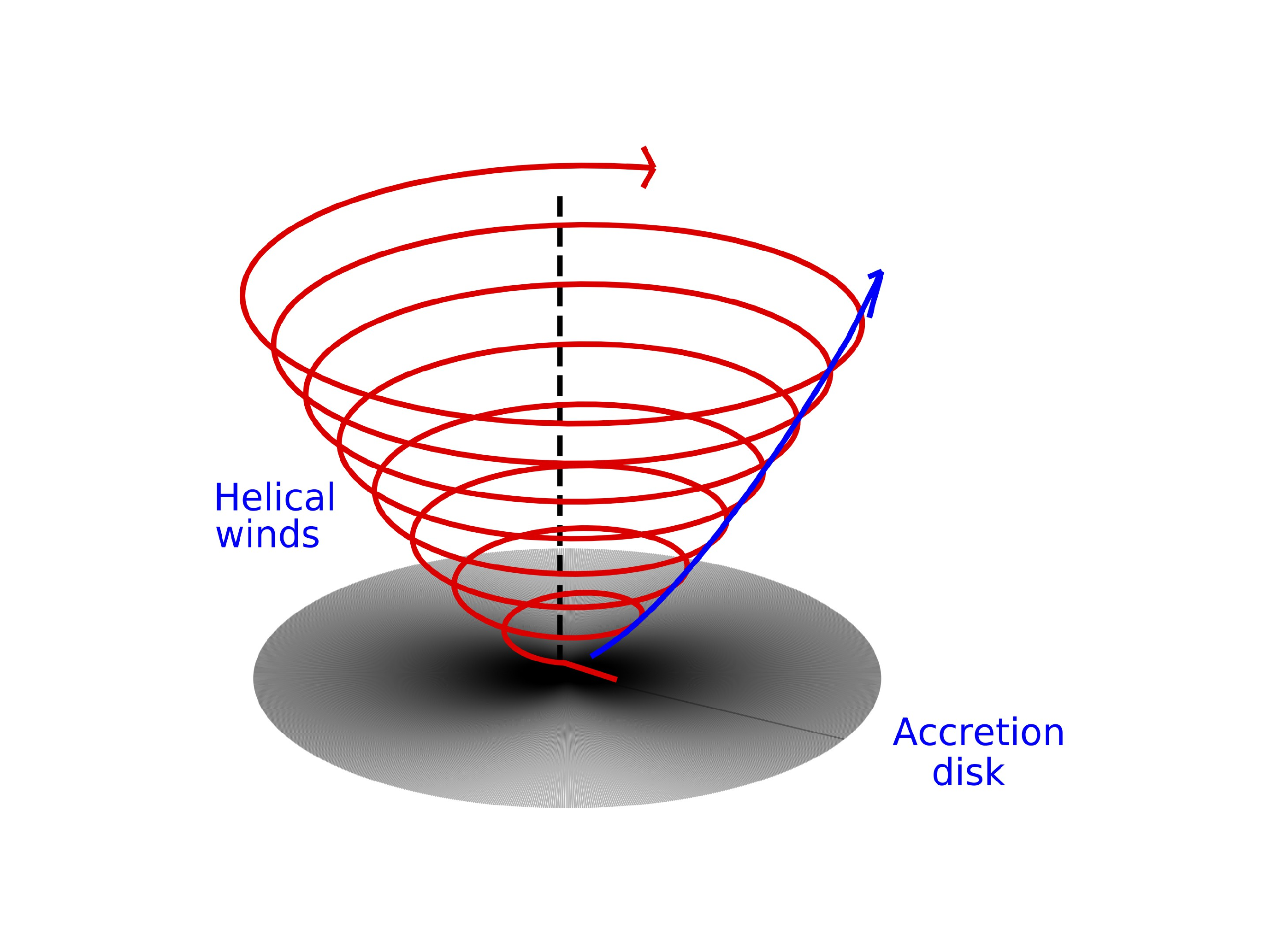}
\includegraphics[width=3.5in]{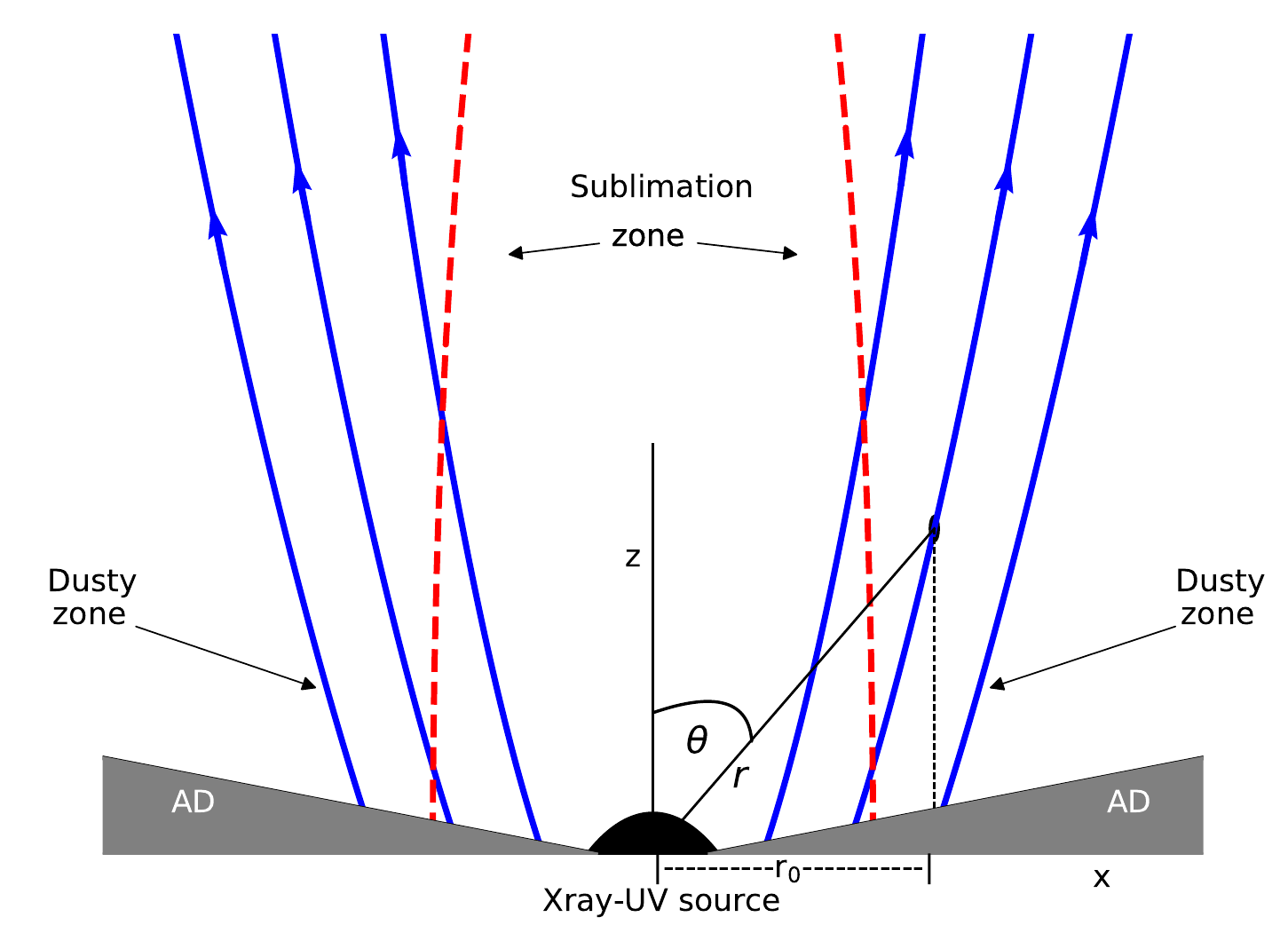}
\caption{\label{geometry}\footnotesize{\textit{Left-panel:} A schematic representation of the helical outflow of the gas from the accretion disk, controlled by the magnetic field of the central BH, is shown in the figure. Assuming azimuthal symmetry, the projection of the motion of the wind in the x-z plane is shown in blue. The arrowheads indicate the direction of flow. For a more accurate presentation of this figure, we refer to \cite{fukumura2018b} (Fig. 3). \textit{Right-panel:} The projection of the accretion disk (AD) is shown along with the outflowing winds (in blue) in the upper-hemisphere. The motion of the outflows along this plane can be represented by parabolic equations of motion. The luminosity of the system is attributed to a central X-ray UV source. The dotted line (in red) represents the sublimation radius and the dust-free zone. }} 
\end{figure*}

\section{The MHD wind model}
\label{sec_phmodel}

The MHD wind model consists of a Keplerian, geometrically thin disk around a supermassive BH of mass $M_{\rm BH}$, threaded by a poloidal magnetic field. The Keplerian disk rotation induces a toroidal field component and an ensuing magneto-centrifugal outflow of the disk plasma \citep{blandford1982}. The axisymmetry of the problem allows the treatment in 2.5D, $i.e.$ the spatial description in the 2D poloidal plane, while keeping track of all three velocity components. 

The mathematical details of the problem can be found in \cite{conto94} and \cite{konigl1994}, while more detailed applications to the ionization structure of the resulting wind and to X-ray observations can be found in \cite{fukumura2010}. The field treads the disk across its entire extent, which spans many decades in radius, and so does the wind's lateral extent. This is consistent with the the X-ray observations that demand such a large lateral wind size \citep{behar2003, holczer2007, fukumura2010}, and justifies the self-similar treatment of the MHD wind models implemented in the above references \citep{conto94, konigl1994}.

Following that, we consider the presence of an underlying, steady-state, self-similar magnetic field driven MHD wind with an $r^{-1}$ density law, along any given line of sight, while also maintaining a steep dependence on the inclination angle. These are consistent with the compilation of density profiles obtained by \cite{behar2009}, while we followed the models of these winds detailed in by \cite{fukumura2010, fukumura2018}. 

\begin{figure}
\vspace*{0.3cm}
\centering
\includegraphics[width=3.5in]{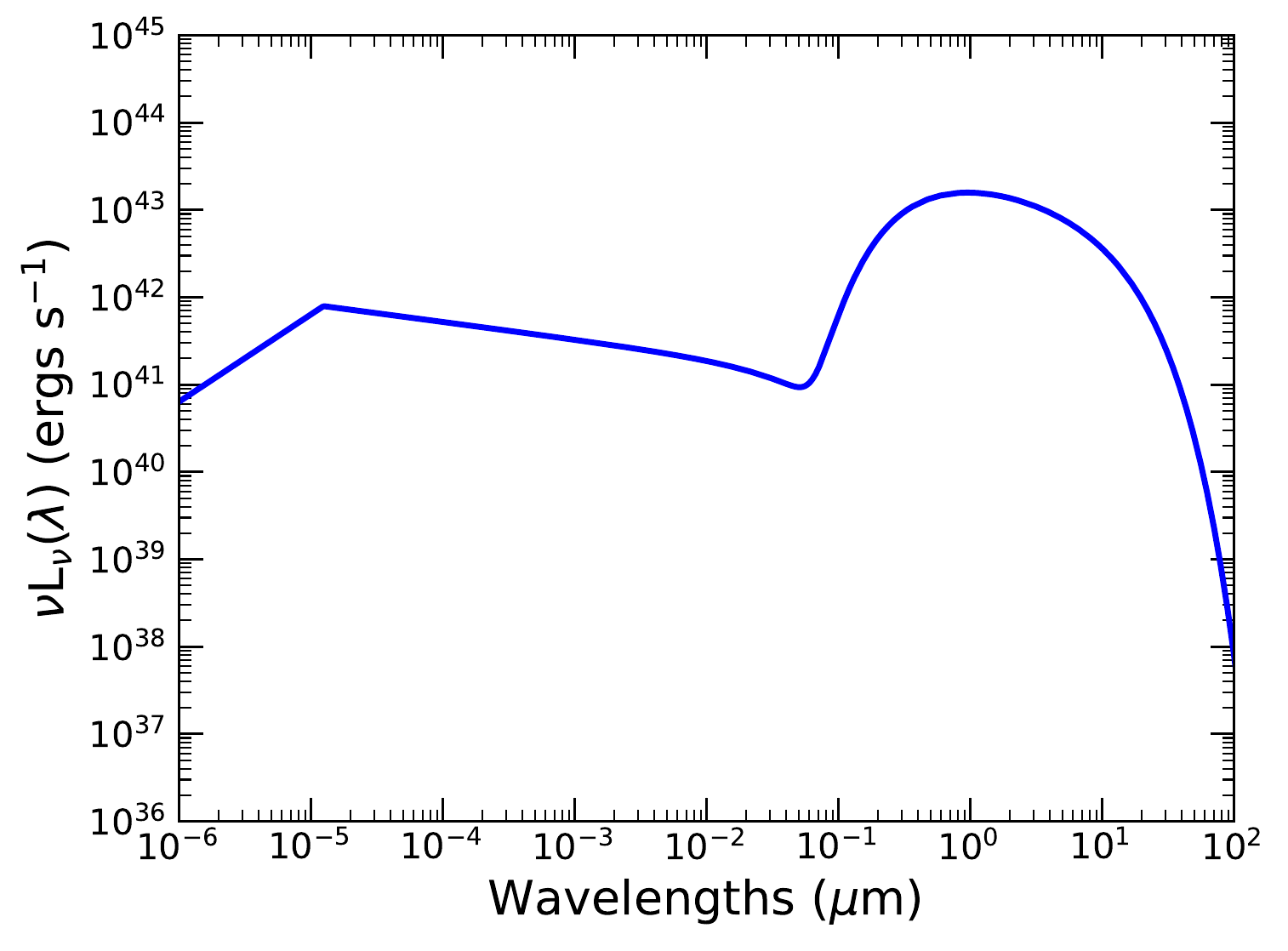}
\caption{\label{source}\footnotesize{The luminosity of the central source $L_{\nu}(\theta)$ is presented for an inclination of $\theta = 75 ^\circ$, defined in Equation (\ref{eq_lum}). The shape of the spectra is determined by $\alpha_\mathrm{{ox}}$ = -1.5. For all the inclination angles, the shape of the spectra remain unchanged whereas the total luminosities vary with $f(\theta)$. The material around the AGN is solely heated by the radiation from the central source.}} 
\end{figure}

These winds can either be dusty or dust free in nature. In this study, we have assumed that the outflows start with an initial dust-free gas of solar composition.  In reality, the actual chemical composition of the outflowing winds is not well known, as it depends on several complex factors related to the nature of the stars around the accretion disk and the underlying physics. We then follow the physical conditions and chemical evolution of the gas along the flow. Our choice of an initially dust free composition sets an upper limit on the production rate of new dust in AGNs. 

The formation of dust can be affected by the possibility of clump formation in the flows, as indicated by similar studies of supernova ejecta by \cite{sar15}. However, in order to keep our models sufficiently simple and manageable, considering the complexity of addressing the chemistry in a 2D geometry, we have assumed a homogeneous flow with $r^{-1}$ density law  over the entire region. The possible effect of clumpiness is discussed in the discussion. 


The outflow in the form of winds from the accretion disk is shown as a schematic diagram in Figure \ref{geometry}. Figure \ref{geometry}(left-panel) shows the upper hemisphere of the accretion disk with the azimuthally symmetric helical winds flowing outwards from the disk-plane. The blue lines in the figure show  the wind-path and flow-line along the 2D xz-plane.   Figure \ref{geometry}(right) further illustrates the flow-lines in 2D geometry with various starting or launch radius on the surface of the accretion disk. The polar coordinates ($r,\theta$) of a given point and the axes are shown on Figure \ref{geometry}(right-panel), where $r$ is the radial distance from the centre of the accretion disk and $\theta$ is the inclination angle with the z-direction. 


As detailed in \cite{conto94}, the poloidal field lines can be approximated in a self-similar model by a set of homothetic parabolas (Fig. 1 right-panel) with the gravitating center (BH) at their focus. The similarity is broken by the BH horizon and the Innermost stable circular orbit (ISCO) at radii a few times the Schwarzschild radius $R\rm_{sch}$ ($R\rm_{sch} \sim 3 \times 10^{5}$~$M_0$ cm; $M_0 = M_{\rm BH}/M_{\odot}$). However, herein we consider radial scales many orders of magnitude larger than $R\rm_{sch}$, warranting the self-similar treatment of the problem. The poloidal flow of the plasma then takes place along each such line. Their detailed shape of these flow lines is given by the solution of the corresponding transfield balance equations (the Grad-Safranov equation). In the present work we refrain from solving explicitly these equations and present only approximate kinematics and geometry, based on the numerical integration of the equations given in \cite{fukumura2010}.

In this 2D geometry the particles at $\theta \gsim 75^{\circ}$ move almost in Keplerian fashion, with their motion becoming progressively poloidal, especially beyond the Alfv\'en point to follow the parabolic magnetic field lines \citep{blandford1977, meier2012}. Placing the central source at the focus (of these parabolas), we consider the following parabolic equation for their shape,

\begin{equation}
\label{motion}
r = r_0/(1-cos\theta)
\end{equation}
where, $r_0$ is the radius where any of these parabolas (magnetic field lines) crosses the accretion disk and denotes the radius at which the wind plasma flow is launched along each such field line.

The luminosity of the central UV-Xray source is given in terms of the Eddington luminosity as, 
%
\begin{equation}
\label{eq_lum}
L(\theta) = \eta \dot m L_\mathrm{{Edd}}(M_8) f(\theta)
\end{equation}
where, $\eta$ ($\sim$ 0.1-1) is the radiative efficiency of an accreting BH, $\dot m$ is the normalized mass accretion rate given by $\dot m = \dot M/\dot M_\mathrm{Edd}$, $L_\mathrm{Edd}$(M$_8$) and  $\dot M_\mathrm{Edd}$ (=$L_\mathrm{Edd}$/$c^2$)  are the Eddington luminosity and Eddington accretion rate for a 10$^8$ \Ms\ BH, $M_8$ is the mass of the central BH normalized by 10$^8$ \Ms\ and $f(\theta)$ is the anisotropic coefficient. For $M_8$ = 1, $L_\mathrm{Edd}$(M$_8$) = 1.3$\times$10$^{46}$ ergs s$^{-1}$. The radiative efficiency was taken as 0.3. The anisotropic coefficient, describing the angular distribution of radiation emitted by a thin disk are given by \citep{netzer1987, stalevski2016}, 

\begin{equation}
\label{anisotropy}
f(\theta) = (1/3) \cos\theta (1 + 2 \cos\theta).
\end{equation}

where $\theta$ is the  inclination angle $\theta$  shown in Figure \ref{geometry}. 

The calculations presented in the following sections assume a BH mass of $10^7$ \Ms\ or $M_8 = 0.1$, the radiative efficiency $\eta \sim 0.3$ and the wind-mass rate is equal to the accretion on the BH, $i.e.$, $\dot m$ = 1 \citep{fukumura2010}. The Eddington luminosity for a 10$^7$ \Ms\ BH is $\sim$ 1.3 $\times$ 10$^{45}$ ergs s$^{-1}$. Because the wind mass flux depends on the radius (generally increases with increasing disk radius), we assumed that its mass flux at the smallest disk radius, $R_{\rm ISCO} \simeq 3 R_{\rm sch}$, is equal to the mass accretion rate onto the BH.

We assume that the material around the AGN is solely heated by the radiation from the central source. The shape of the spectrum employed in our calculations is given in Fig. \ref{source}, properly normalized for the inclination angle of 75$^{\circ}$. The AGN spectra are usually characterized by a power law with an index $\alpha_\mathrm{{OX}}$  between 2100 \AA\ and 2 keV ($\sim 0.2$ and 6$\times$10$^{-4}$ $\mu$m).  The values of $\alpha_\mathrm{{OX}}$ range between $-1$ and $-2$; the spectrum presented in Fig. \ref{source} has $\alpha_\mathrm{{OX}} = -1.5$. For other inclination angles, the same shape of the source spectra is retained and the effective luminosity is calculated using $L(\theta)$ from Equations \ref{eq_lum} and \ref{anisotropy}. 


\begin{figure*}
\vspace*{0.3cm}
\centering
\includegraphics[width=3.55in]{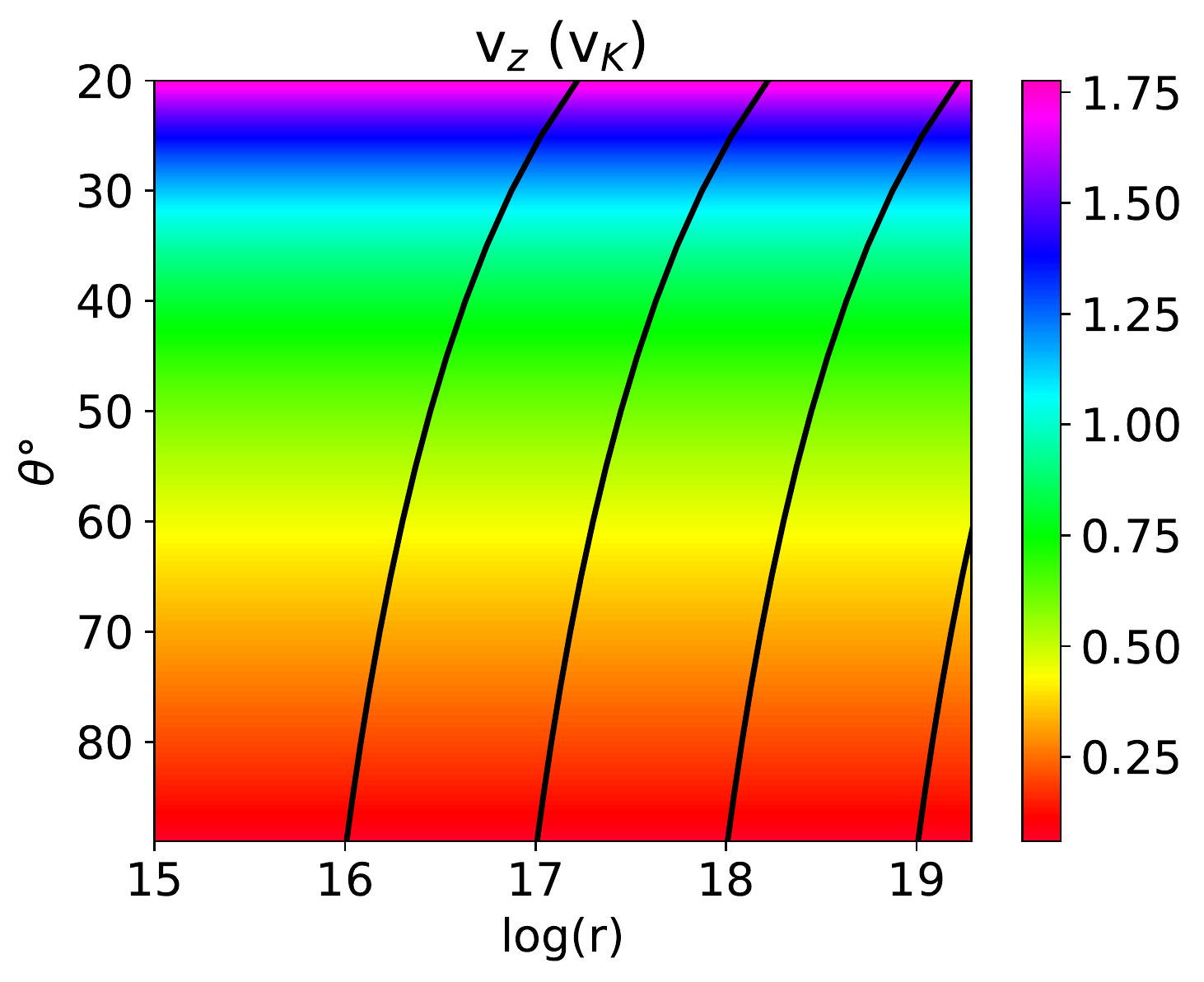}
\includegraphics[width=3.45in]{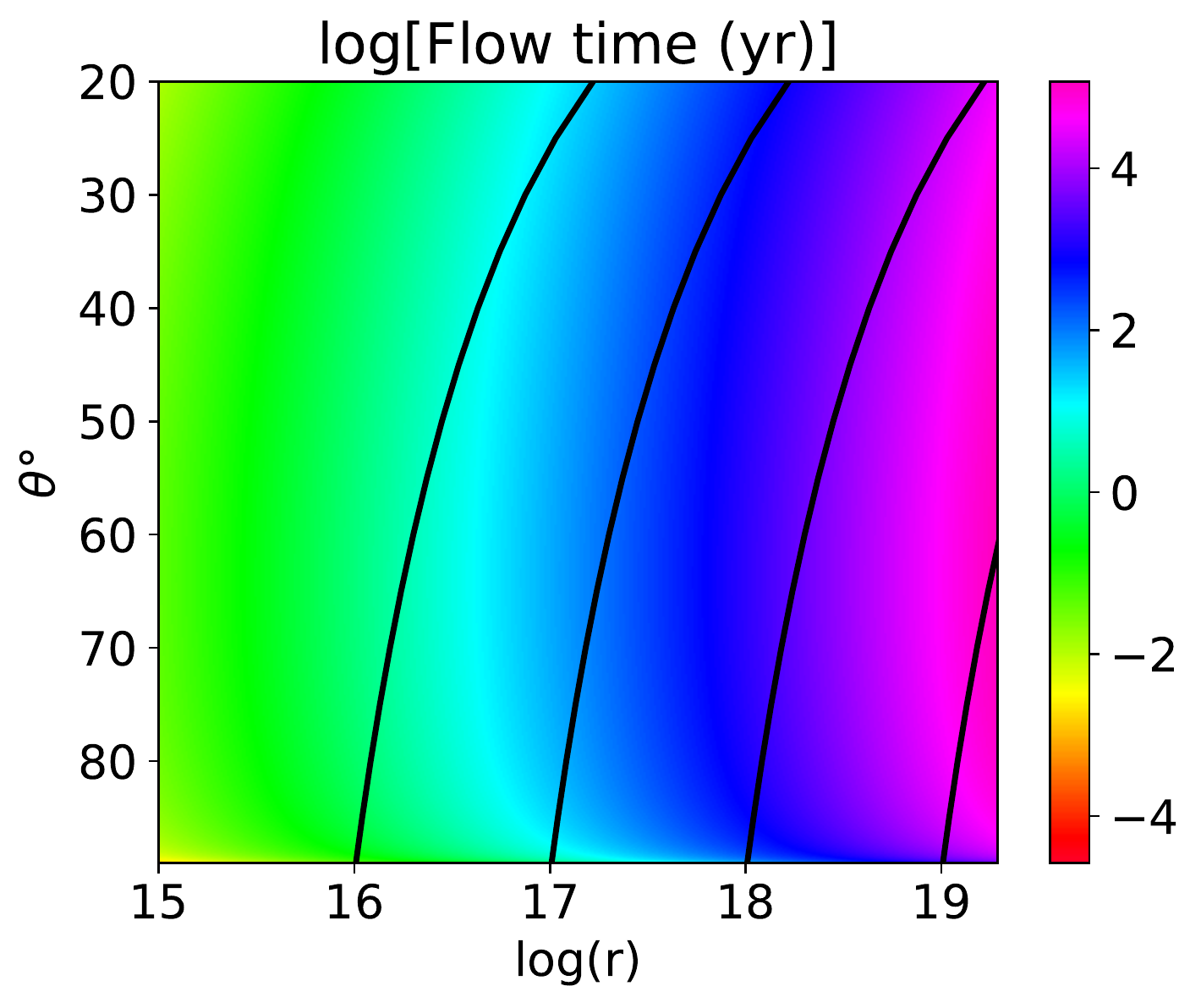}
\includegraphics[width=3.45in]{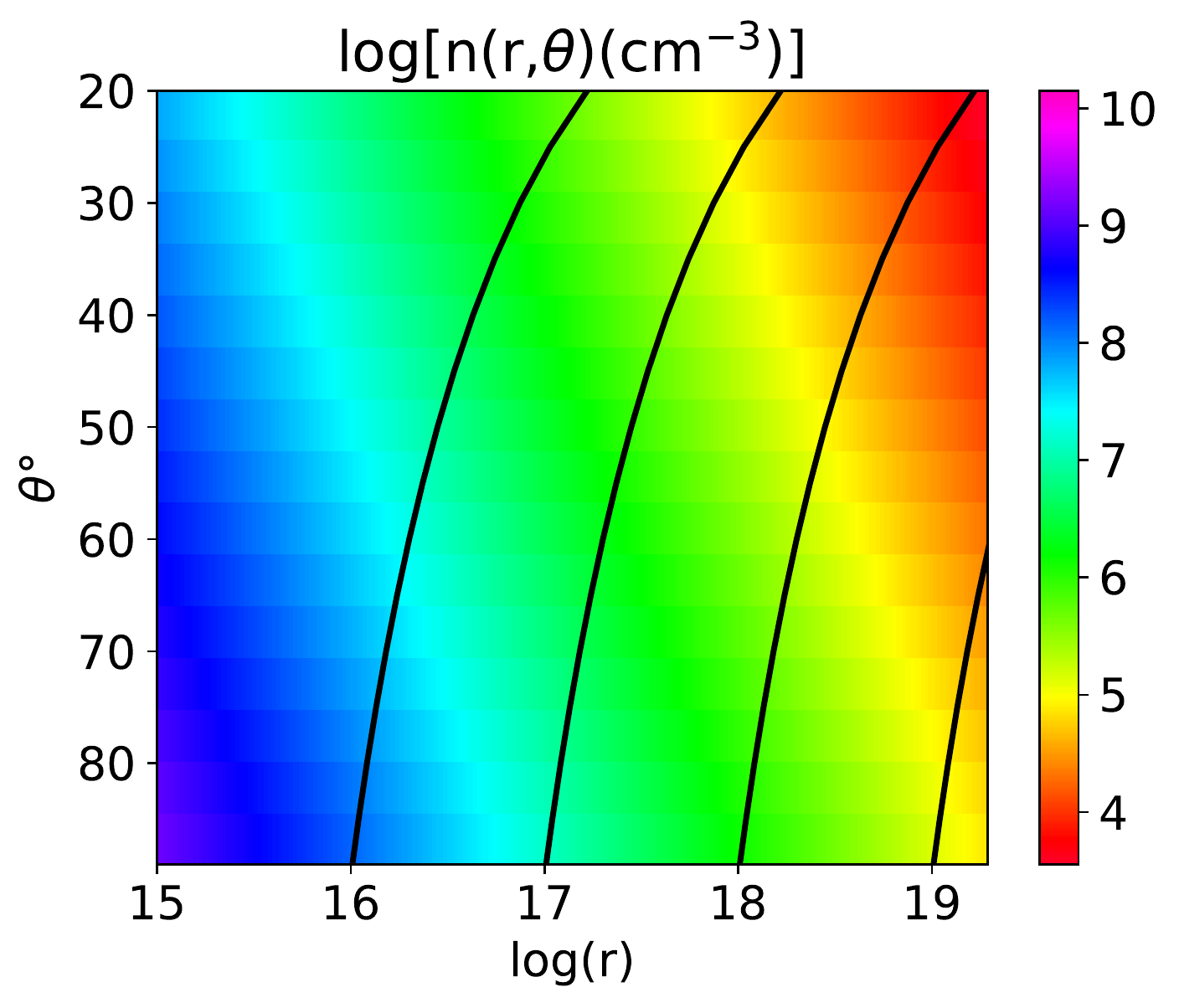}
\includegraphics[width=3.45in]{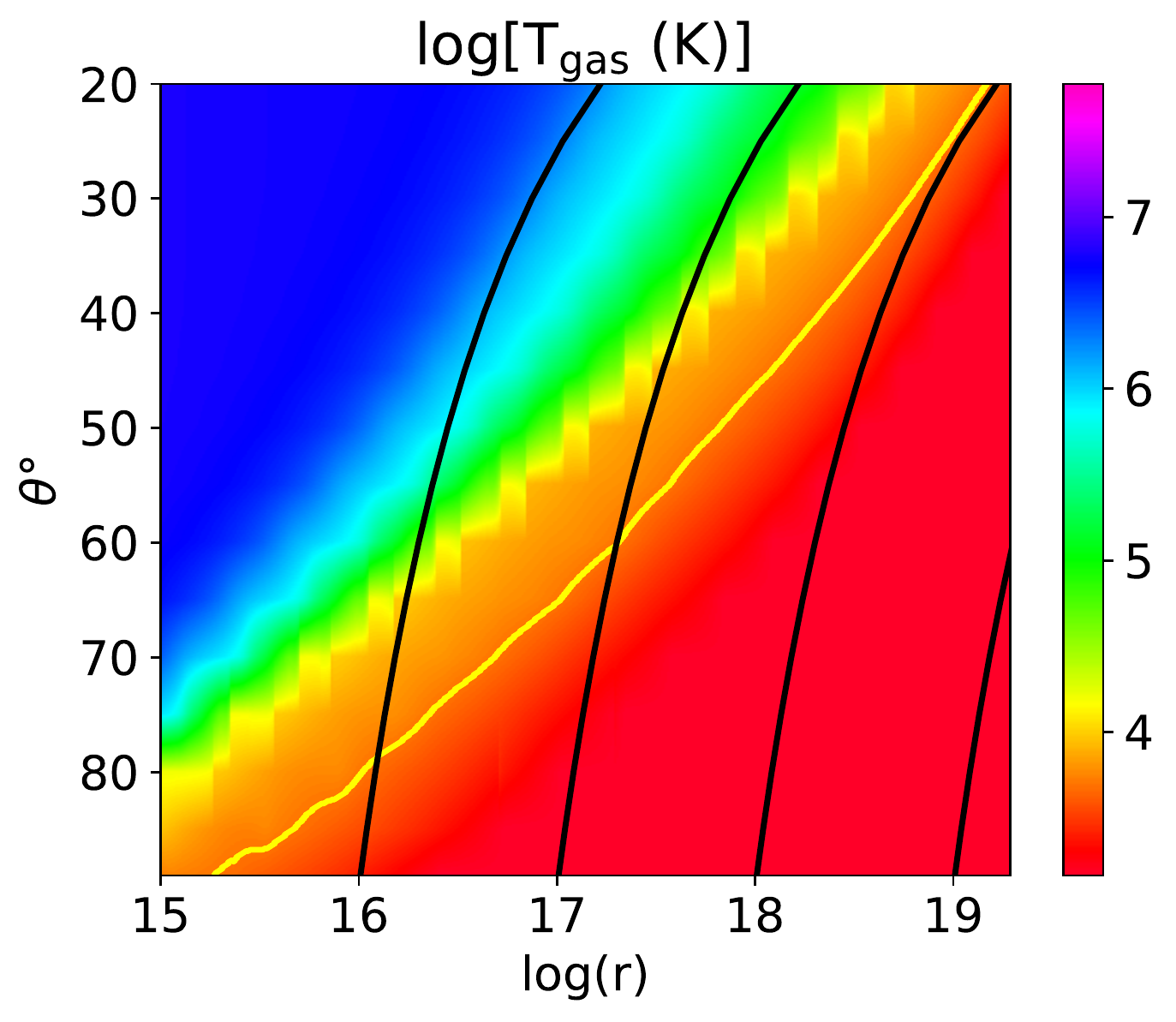}
\caption{\label{fig_v_t_T_n}\footnotesize{The colormap of physical quantities in polar ($r, \theta$) coordinates are shown. The lines (in black) on the figure represents the parabolic flow-lines with various starting/launch radius on the accretion disk, $r_0$. The accretion disk is located along the $\theta = 90 ^\circ$ plane, $i.e.$, the lower-horizontal axis of the figures. \textit{Top-left:}  Gas velocities in the z-direction in terms of the Keplerian velocities (V$_K$) are presented (Equation (\ref{eq_vz})). \textit{Top-right:} The time (in years) taken by the gas to reach a given distance on the $r$-$\theta$ plane is presented (Equation (\ref{eq_time})). \textit{Bottom-left:} The distribution of gas gas density $n(r,\theta)$ in polar coordinate is presented, assuming solar composition of the gas. \textit{Bottom-right:} The colormap shows the gas temperature in polar ($r, \theta$) coordinates, assuming solar composition of the gas. The line in yellow demarcates the region lower than 4000 K, in the $r, \theta$ plane, where formation of molecular clusters and seeds are supported. Importantly to note however, the formation of dust grains from these seeds happens only when the temperature is lower than 2000 K.  }} 
\end{figure*}


Our region of interest encompasses a cylindrical volume above and below the accretion disk bound by the gravitational sphere of influence of the central BH. The radius of influence ($R_\mathrm{infl}$) is approximately the distance on the accretion disk where the total mass of the stars equals the mass of the central BH \citep{thomas2016} and is determined by the velocity of dispersion ($\sigma$) of the stars on the disk \citep{gebhardt2000, merritt2000, mcconnell2013}. It is given by the equation, 

\begin{equation}
\label{infl_radius}
R_\mathrm{infl} \sim R_{\mathrm{sch}} (c/\sigma)^2
\end{equation}
where, $R_\mathrm{sch}$ is the Schwarzschild radius of the central BH and $c$ is the speed of light. We take $\sigma$ = 200 km~s$^{-1}$ for this study \citep{forbes1999}. An inclination of $\theta = 20 ^\circ$ is taken as the upper edge of the cylindrical volume as the wind have already crossed the Alfv\'en point at the altitude \citep{fukumura2010}. Beyond the corresponding latitude, its $v_z$ becomes super-Keplerian  \citep{fukumura2010}, and as such, we do not expect dust formation beyond this point. For a 10$^7$ \Ms\ BH,  $R_\mathrm{sch}$ $\sim$ 2.94 $\times$ 10$^{12}$ cm. Therefore, $R_\mathrm{infl}$ $\sim$ $r_\mathrm{0,max}$ $\sim$ 6.6 $\times$ 10$^{18}$ cm. The outer radius of this cylindrical volume is $R_\mathrm{max}$ $\sim$ $R_\mathrm{out}$ $\sim$ $r_\mathrm{0,max}$/$sin\theta_\mathrm{min}$ = 1.9 $\times$ 10$^{19}$ cm.

The fluid velocity at low latitudes (close to the plane of the disk) is mainly in the toroidal direction at the Keplerian value,  with the toroidal magnetic and velocity fields assuming a corkscrew geometry as shown in Figure \ref{geometry}. At higher latitudes ($\theta$ small), the flow passes beyond the Alfv\'en point and the magnetic field cannot enforce corotation. The magnetic field lines bend backward and angular momentum conservation forces the toroidal velocity (and field) to fall-off like $1/r \sin \theta$ while the dominant fluid velocity is in the poloidal direction \citep[see e.g][]{fukumura2010}. 

For the velocity of the plasma in the $z$-direction, we use an approximation of the expression derived numerically in \cite{fukumura2010}, valid up to the Alfv\'en radius, i.e. to an inclination angle of $\theta \simeq 20^{\circ}$, an approximation sufficient for our purposes since the plasma is much too dilute at larger latitudes to be of any significance in the formation of dust. Therefore, we fit the poloidal velocities calculated by \cite{fukumura2010} and estimate the velocity in polar direction ($v_z$) at all points. Considering the helical motion, the velocity of Keplerian rotation (V$_K$) is given by $(GM_\mathrm{BH}/x)^{1/2}$. The best fit case of $v_z$ at a given coordinate ($r_1,\theta_1$) is presented as, 

\begin{equation}
\begin{split}
\label{eq_vz}
& v_z =  \beta  \Big(\frac{z}{r_0}\Big)^{p} V_K, \ \ \ \ \beta = 0.45, \ \ \  p = 0.5 \\
& v_z(r_1,\theta_1)=  \beta \Big(\frac{GM_\mathrm{BH}}{r_0} \frac{cos\theta_1}{sin\theta_1}\Big)^{1/2}
\end{split}
\end{equation}
where, $\beta$ and $p$ are fitting parameters, $r_1$ = $r_0/(1-cos\theta_1)$. The equation of the flow (and poloidal magnetic field) along the line (Eqn. \ref{motion}), combined with the expression for the $z$-component of the velocity flow (Eqn. \ref{eq_vz}) below, can determine the motion of the gas in the direction, most importantly the time spent at each latitude, as this should be sufficiently long for the formation of dust.

Figure \ref{fig_v_t_T_n} shows the different physical conditions in the flow as a function of the polar coordinates. 
 The top left panel shows the variation of $v_z$ as a function of the Keplerian velocities (V$_K$) in the polar coordinate ($r\theta$) plane.  As seen in the figure, initially when a parcel of gas is closer to the plane of the accretion disk, the polar velocities are small. As the gas parcel flows further away (smaller $\theta$), the velocities continue to increase, until it crosses the Alfv\'en radius. 
   
The top right panel show the flow time along the poloidal magnetic field lines. The flow-time is the time ($t$) taken by the wind to reach a given point ($r_1,\theta_1$) on the $r\theta$-plane. It is calculated using the equations of motion, given by Eqn. \ref{motion} and \ref{eq_vz}, as shown in the equation below.  Dust formation requires that the timescales for chemical reactions and nucleation processes be shorter than the flow time. 

\begin{equation}
\begin{split}
\label{eq_time}
& z = \frac{r_0 \ cos\theta}{1-cos\theta},  \ \ \ \ \frac{\mathrm{d}z}{d\theta} = -\frac{r_0sin\theta}{(1-cos\theta)^2} \\
& \frac{\mathrm{d}z}{dt} = \frac{\mathrm{d}z}{d\theta}\frac{\mathrm{d}\theta}{dt} =  -\frac{r_0sin\theta}{(1-cos\theta)^2} \frac{\mathrm{d}\theta}{dt} = \beta \Big(\frac{GM_\mathrm{BH}}{r_0} \frac{cos\theta}{sin\theta}\Big)^{\frac{1}{2}} \\
& dt = - \frac{r_0^{3/2}}{\beta (GM_\mathrm{BH})^{1/2}} \frac{\ sin^{3/2}\theta}{(1-cos\theta)^{2} cos^{1/2}\theta}  d\theta \\
& t(r_1,\theta_1) = - \int^{\theta_1}_{\theta > \pi/2} \frac{r_0^{3/2}}{\beta (GM_\mathrm{BH})^{1/2}} \frac{\ sin^{3/2}\theta}{(1-cos\theta)^{2} cos^{1/2}\theta}  d\theta \\
\end{split}
\end{equation}

 The flow-time is proportional to the launch radius $r_0$ to its 3/2 power, as befits the free-fall time in a gravitational field. Therefore, at a larger launch radius $r_0$, the local wind velocities are slower. To give an example, outflows starting from the inner regions ($r_0$ $\sim$ 10$^{16}$ cm) of the accretion disk reaches an inclination angle of $\theta = 20 ^\circ$ within an year, where as in the outer regions ($r_0$ $\sim$ 10$^{18}$ cm) reaches at the same height in more than 10$^4$ years. 

The bottom left panel shows the gas density in the $r-\theta$ plane. For the gas density, including its angular distribution, we have employed a fit to the expression given in \cite{fukumura2010}(Fig. 2a), who integrated the wind equations choosing the poloidal current in a way that the radial density dependence be proportional to $r^{-1}$, in agreement with the observations of \cite{holczer2007}. 
As in \cite{fukumura2010}, we normalize the radius $r$ in units of the Schwarzschild radius, $R_{\rm sch} = 3 \times 10^{13} M_8$ cm, where $M_8$ is BH mass normalized to $10^8$ solar masses, and the dimensionless unit of length $X = r/R_{\rm sch}$.
For the angular distribution we simply use a fitting formula to the results of \cite{fukumura2010} (fig. 2a). The density is normalized so that the Thomson depth over the BH Schwarzschild radius be equal to the dimensionless accretion rate $\dot m$, i.e. $n_M \, \sigma_{\rm T} \, R_{\mathrm{sch}} = \dot m$ or $n_M \simeq n_0 (\dot m/M_8)$, where $n_0$ = $5 \times 10^{10}$. So the density has the form,
 
\begin{equation}
\label{eq_density}
n(r,\theta) = n_0 \frac{\dot m}{M_8} X^{-1} e^{\frac{5}{2}(\theta - \frac{\pi}{2})} 
\end{equation}
As seen in the figure, the gas densities are higher at smaller radii and large inclinations ($r$ small, $\theta$ large). Therefore, the density of the gas in the outflows gradually decrease along its path, as the wind moves outwards and towards higher latitudes (low $\theta$).

In this study, $M\rm_{BH}$ is 10$^7$ \Ms, $M_8$ is 0.1, $R_\mathrm{sch}$ $\sim$ 2.94 $\times$ 10$^{12}$ cm and $\dot m$ = 1 (explained with Equation (\ref{eq_lum})). Using Equation (\ref{motion}), we have, 
\begin{equation}
\label{eq_density_easy}
n(r,\theta) = \frac{C_0}{r_0} (1-cos\theta)e^{\frac{5}{2}(\theta - \frac{\pi}{2})}, \ \  C_0 =  \frac{n_0 \dot m R\rm_{sch}}{M_8} 
\end{equation}

 The bottom right panel shows the gas temperature distribution derived in Section~\ref{sec_gasT}  below.

\begin{figure}
\vspace*{0.3cm}
\centering
\includegraphics[width=3.5in]{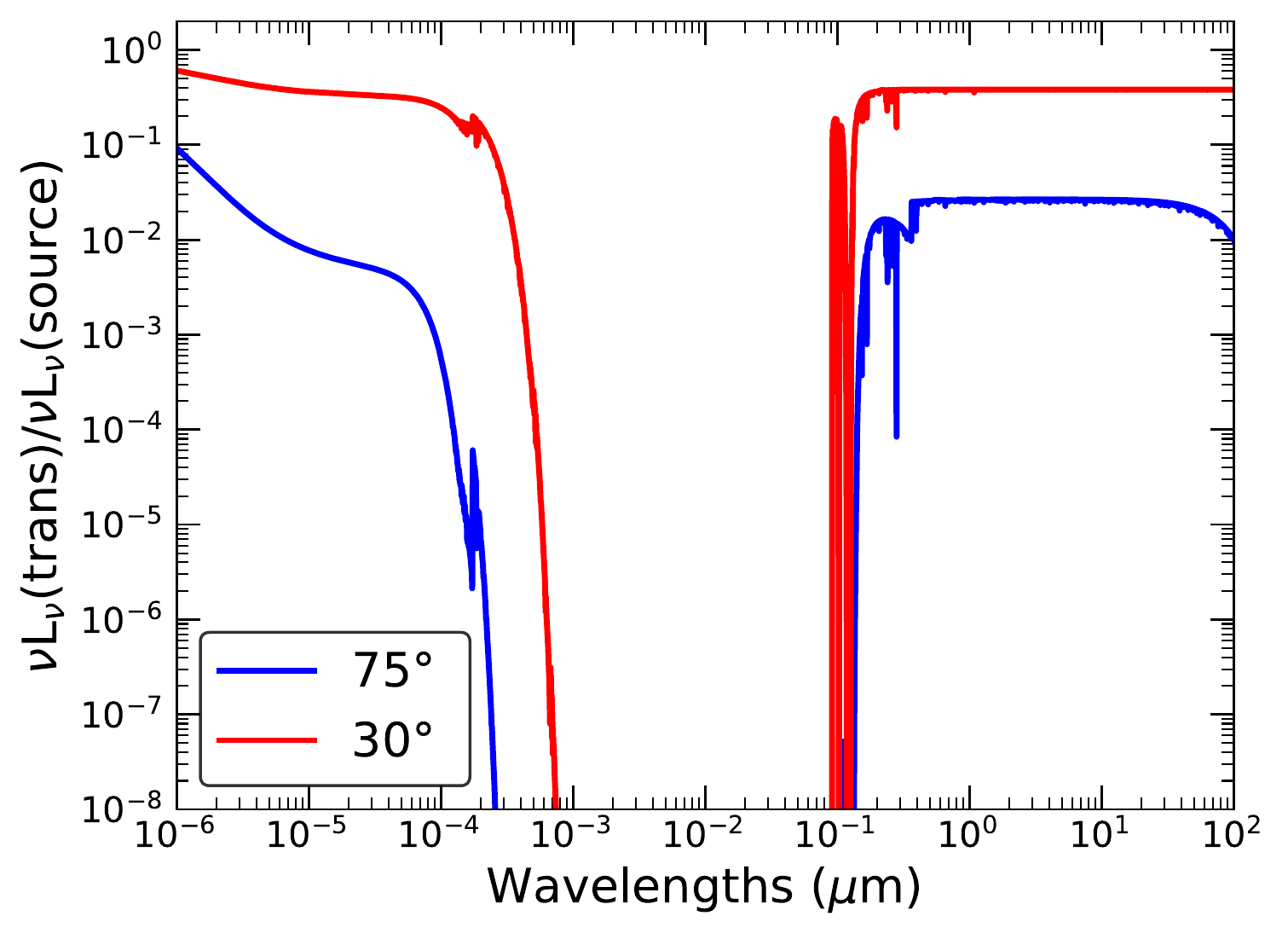}
\caption{\label{fig_lum_temp_gas}\footnotesize{The figure shows the ratio of the luminosities transmitted through the column of outflowing gas from the accretion disk, compared to the source luminosities, for two inclinations, $\theta = 30^\circ, 75^\circ$. Owing to the larger column densities and high densities along high inclination angles, a larger fraction of the source luminosity is absorbed by the gas.}} 
\end{figure}

\section{The thermodynamic conditions for the formation of dust}
\label{sec_results}

\subsection{The gas temperature}
\label{sec_gasT}

The formation of dust proceeds through a series of chemical processes, such as chemical reactions, nucleation, coagulation and gas accretion onto dust, which involves atoms, ions, molecules, stable clusters (seeds) and solid grains \citep{sar15, gob16}. The collective process requires high gas densities in order to ensure adequate number of collisions within a flow timescale. It also requires moderately high gas temperatures to sustain higher thermal velocities needed to overcome energy barriers where necessary. 


The gas temperature is determined by the balance between its radiative and adiabatic expansion cooling and its heating by the illumination by the central source.
We have calculated the gas temperature employing heat balance as a function of inclination angle.  The central source with luminosity $L(\theta)$, given by Equation (\ref{eq_lum}), and the shape of the spectra shown in Figure \ref{source}, is taken as the heating source. The gas density of the medium is taken as $n(r,\theta)$ from Equation (\ref{eq_density}). The radiative transfer through various inclination planes is computed using the spectral synthesis code CLOUDY \citep{fer98} to determine the local absorption coefficient and the spectral flux at each point of the flow. We assume a gas of solar composition \citep{grevesse1998}.

Figure \ref{fig_lum_temp_gas} (left) presents the ratio of the transmitted luminosity to the intrinsic source luminosity, at the outer radius of the cylindrical volume of gas, $R_\mathrm{out}$ (defined in Section \ref{sec_phmodel}), for inclination angles $\theta = 30^\circ, 75^\circ$. For both the cases, the soft X-rays ($\sim$10$^{-3}$-10$^{-1}$ $\mu$m) are heavily attenuated by the gas, while the hard X-rays ($\lambda$$<$10$^{-3}$ $\mu$m) and non-ionizing radiations at wavelengths larger than 0.1 $\mu$m are only partially absorbed. It is evident that at $\theta = 75^\circ$, the attenuation of the source radiation is much higher compared to the attenuation at $\theta = 30^\circ$. This is mainly because of a larger gas densities and larger total column density along higher inclination angles (larger $\theta$), as shown in Figure \ref{fig_v_t_T_n} (bottom-left). The rate of radiative cooling was found to be dominant over adiabatic cooling rates for all densities in our region of interest.  

Figure \ref{fig_v_t_T_n} (bottom-right) shows the resultant distribution of gas temperatures in the $r$-$\theta$ plane. The flow-lines shown (in black) on the figure help us to read the gas temperatures along the path of the flow. The gas temperatures range between 10$^3$ to 10$^7$~K within this region of interest. The temperatures along lower inclination directions ($\theta$ small) are found to be much higher compared to the inclination planes which are closer to the accretion disk ($\theta$ large, close to the disk). Moreover, the figure also confirms that the gas is cooler towards the outer radii because of the decline of the incident flux with radius. Radiative cooling rate of a parcel of gas in proportional to the square of gas density \citep{dal72}. Due to high densities close to the accretion disk ($\theta$ large), the cooling is faster, which leads to a lower temperature of the gas at equilibrium. Moreover, the luminosities are also lower along high inclinations (Eqn. \ref{eq_lum}), which aid the lowering of gas temperatures.


As a parcel of gas flows along the flow-line, it moves toward larger $r$ and smaller $\theta$. Along its path, its temperature is determined by radiative and adiabatic cooling and heating by the continuum source. Because adiabatic cooling is always subdominant in the cases of interest, the gas temperature is determined by its photoionization parameter $\xi = L/n(r) r^2$, provided to us by the code CLOUDY \citep{fer98} once the value of $\xi$ is given. Because the particle density drops faster than the photon density with latitude, the gas temperature increases with decreasing $\theta$.

Formation and nucleation of molecules are the first steps towards forming dust grains \citep{cherchneff2008, cherchneff2009, cherchneff2013b}. Many of the molecular reactions are neutral-neutral processes and they are only efficient when the ambient gas temperatures fall below $\sim$ 4000~K \citep{sar13}. In Figure \ref{fig_v_t_T_n} (bottom-right) we have demarcated the region (yellow line), below which the temperature is lower than the threshold temperature. We call the cooler zone below the demarcation boundary as the nucleation zone. The figure indicates that the volume of gas in the nucleation zone is larger towards outer radii and the time spent by the winds in this zone is also longer. The formation of molecules and stable clusters, that act as the seed/monomers to dust grains, takes place in this zone. Once the winds move out of the nucleation zone, formation of molecules and clusters is no more feasible.

\begin{figure}
\vspace*{0.3cm}
\centering
\includegraphics[width=3.5in]{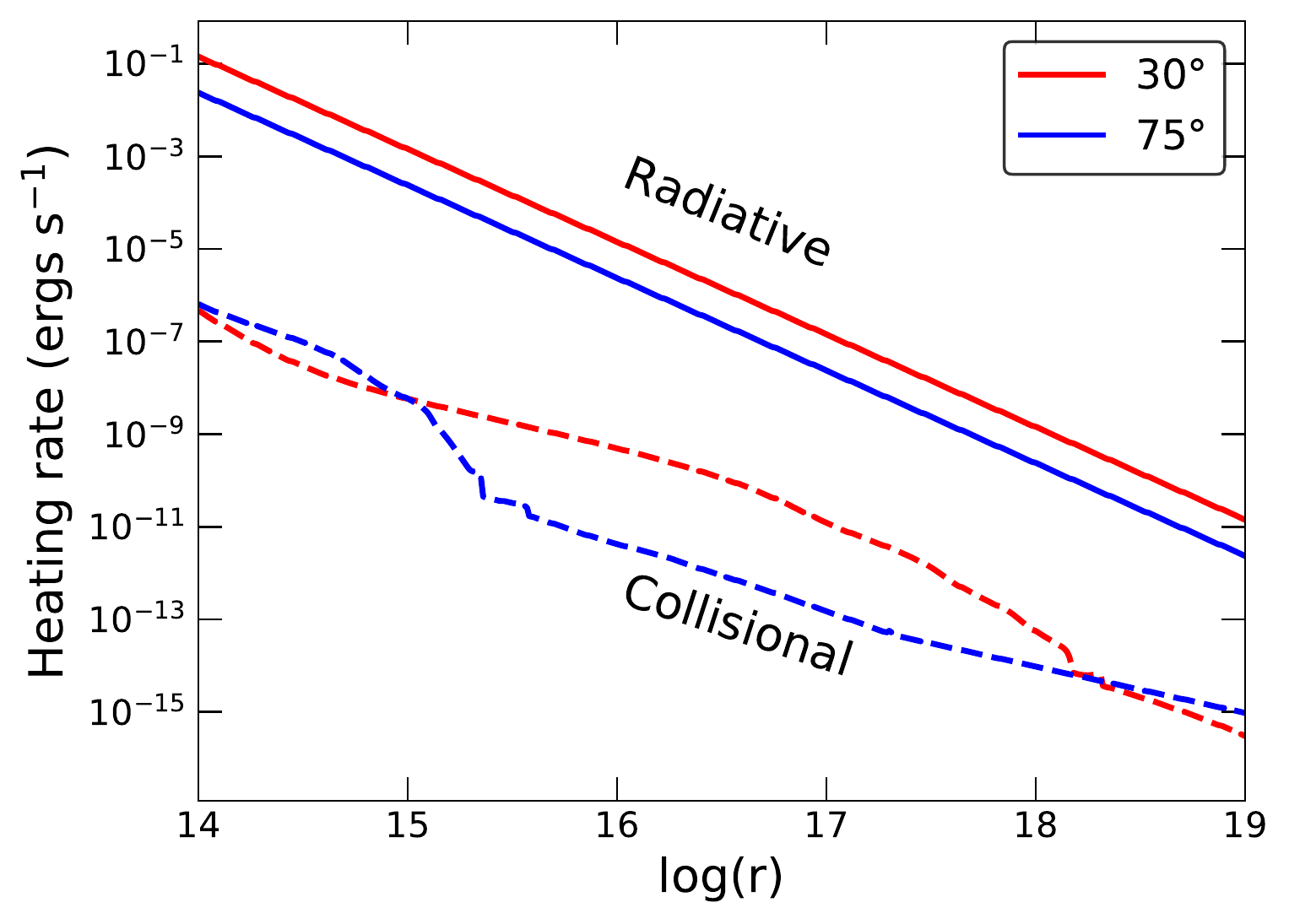}
\caption{\label{fig_heatingrate}\footnotesize{The heating rates of a 10~\AA\ silicate grain as a function of the radial distance is shown for two inclination angles, $\theta = 30^\circ, 75^\circ$, subjected to radiative heating by the central source and collisional heating by the ambient gas. For both the cases the figure shows that, owing to the low densities of the gas, the collisional heating is found to be negligible, compared to the rate of radiative energy absorption. }} 
\end{figure}

\begin{figure}
\vspace*{0.3cm}
\centering
\includegraphics[width=3.5in]{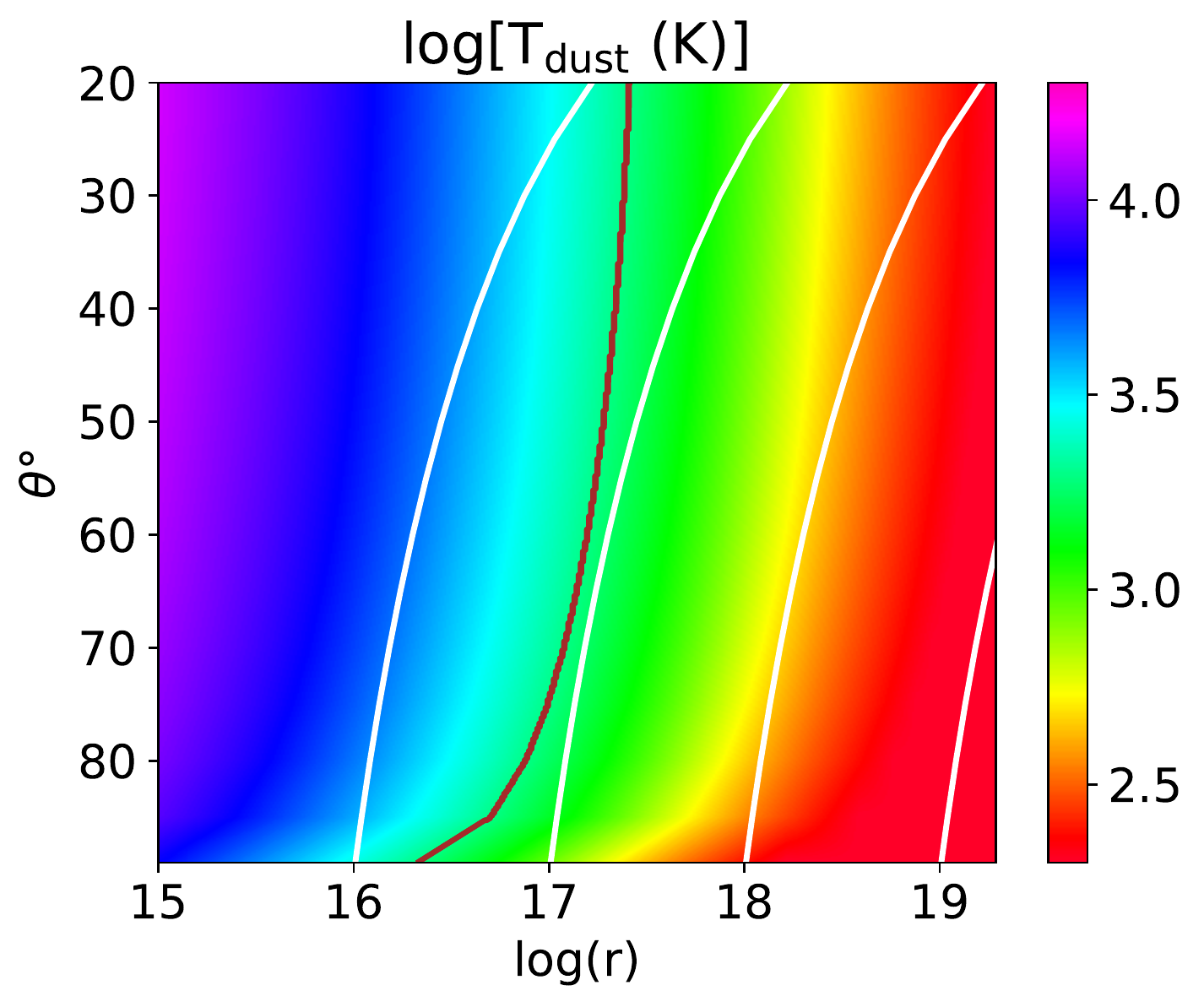}
\caption{\label{fig_dustT}\footnotesize{The colormap shows the dust temperature in polar ($r, \theta$) coordinates for a 10~\AA\ grain of silicate. The lines in white on the figure represents the parabolic flow-lines with various starting/launch radius on the accretion disk, $r_0$. The brown line demarcates the region where dust temperatures are lower than the the sublimation temperature, taken to be $\sim$ 2000 K. Hence, the region in the left-side of the line is the sublimation zone, which is free of dust grains. }} 
\end{figure}

\subsection{The dust temperature} 
\label{sec_dustT}

Formation of dust grains proceeds through the condensation of stable molecular clusters. For a gas of solar composition, silicates are expected to be the primary dust component. We consider a 10~\AA\ grain of astronomical silicate as the prototype of the smallest dust grain. Once formed, these  grains grow by coagulation and accretion thereafter. 

Dust grains are radiatively heated by the incident radiation from the central source and collisionally heated by the ambient gas. The collisional heating rates are estimated as a function of gas densities and gas temperatures using the analysis by \cite{dwe87} and \cite{hol79}, which are recently applied by \cite{sarangi2018}. The collisional heating rate is compared to the rate of radiative heating by the central source. Figure \ref{fig_heatingrate} shows the rate of heating of a 10~\AA\ silicate grain as a function of radial distance from the source, for two inclination angles. It is clear from the figure that for both the cases radiation is the primary heating source, as the collisional heating rates are much lower. In such environments, gas and dust temperatures are individually controlled by radiative heating. 
 At the high densities the gas and dust temperatures are coupled but at different values. 

The dust temperatures are calculated using the heating and cooling balance of the individual grains \citep{dwe87}, subjected to the radiation from the central source. 
Most of the incident radiation is found to be in the range of 0.1-1 $\mu$m wavelength, as evident in Figures \ref{source} and \ref{fig_lum_temp_gas}. The mass absorption coefficient, $\kappa(a_\mathrm{i},\lambda)$, of astronomical silicates is taken from the study by \cite{wei01}. The sublimation temperature is taken as $\sim$ 2000~K. Figure \ref{fig_dustT} shows the distribution of dust temperature in polar coordinates for a hypothetical 10~\AA\ grain. The sublimation zone (T$_\mathrm{dust}$ $\geq$ 2000 K) and the condensation zone (T$_\mathrm{dust}$ $<$ 2000 K) are separated by the demarcation line (in brown). Newly formed grains as well as pre-existing dust can exist when the gas flows through the condensation zone (on the right of the line).

\begin{figure}
\vspace*{0.3cm}
\centering
\includegraphics[width=3.5in]{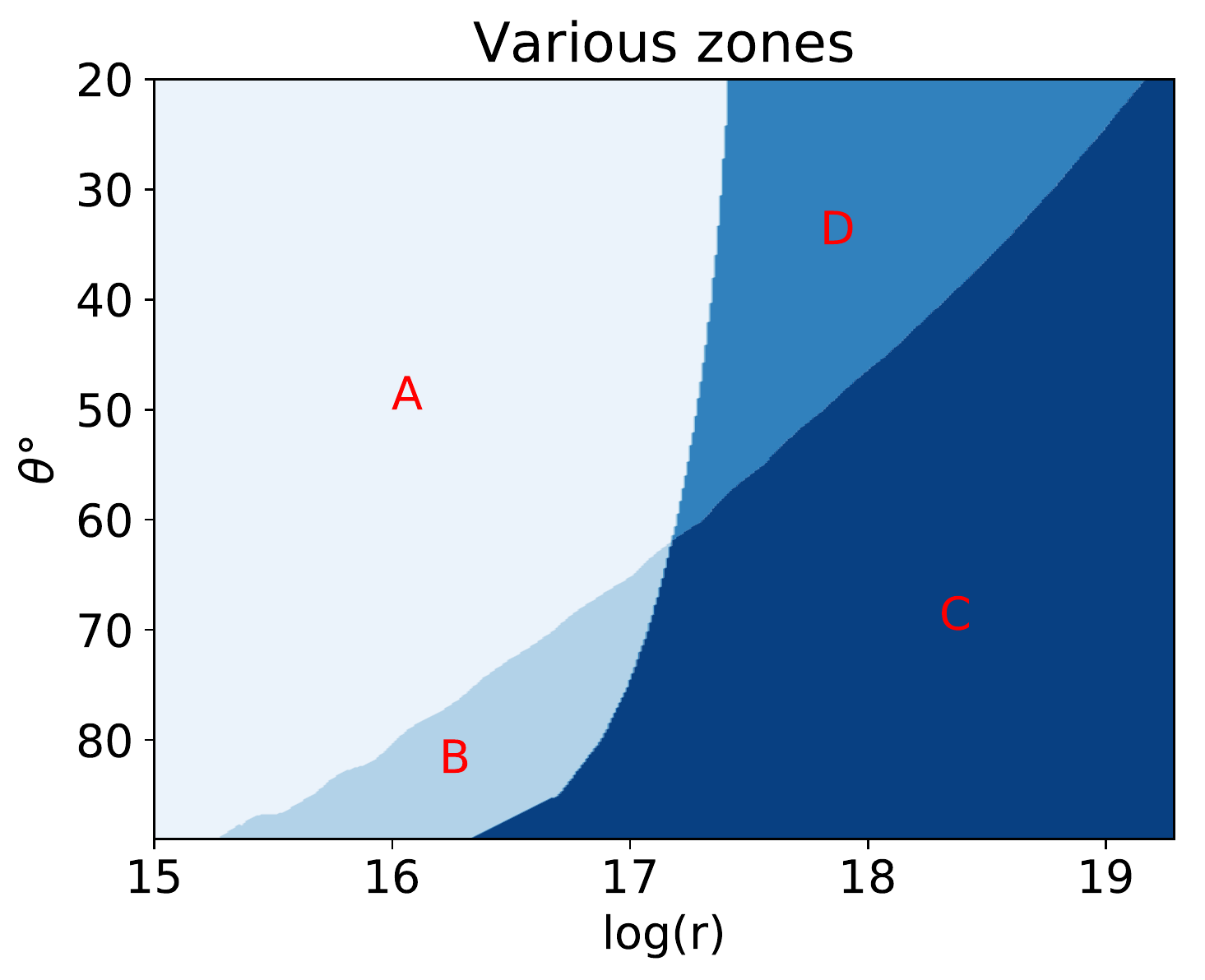}
\caption{\label{fig_zones}\footnotesize{The figure shows the colormap of different regions in the outflows which are are characterized by the following features: (A) region where seeds do not form and dust grains do not survive, (B) region where new seeds form but do not grow (C) region where new seeds form and also simultaneously grow through accretion and coagulation, (D) region where seeds grow via accretion and coagulation but new seeds do not form in the gas}} 
\end{figure}

\begin{table}
\centering
\caption{Various zones in the  $r$-$\theta$ plane}
\label{table_zones} 
\begin{tabular}{cccc}
\hline \hline
 Zone & Nucleation & Condensation & Sublimation \\
 \hline
A &  \ding{55} & \ding{55} &  \ding{52} \\
B  & \ding{52} & \ding{55} & \ding{52}  \\
C  & \ding{52} & \ding{52} & \ding{55} \\ 
D  &\ding{55}  & \ding{52} & \ding{55} \\ 
\hline
\end{tabular}
\end{table}

\subsection{Regions where molecules and dust can form and survive} 
\label{sec_regions}

Based on the temperatures of the gas and dust, various zones of the outflowing winds can be categorized in light of their role in molecule and dust synthesis. The combined contours defined from Figure~\ref{fig_v_t_T_n} and \ref{fig_dustT} mark the boundaries of these zones. Figure \ref{fig_zones} shows the regions labeled by A, B, C and D on the $r$-$\theta$ plane, which are characterized as nucleation, condensation and sublimation zones in Table~\ref{table_zones}. 

Region A comprises of gas that is inside the sublimation zone and outside the nucleation zone. In this region, the hypothetical dust temperatures are too high to sustain the stable grains and also the ionization state of the gas prevents the synthesis of stable seeds, which if formed quickly evaporate because of their high temperature. Region B is defined by the part of the nucleation zone which is inside the sublimation area. Here, gas phase reactions can proceed towards forming stable seeds yet formation of dust from these seeds will not commence when the wind flows through this zone. Region C is the overlapping area of the nucleation and the condensation zone. This the most suitable zone for molecules and dust formation. In region D, temperatures are too high to support the formation of new seeds, however the already formed seeds can grow by coagulation and by accretion of refractory metals. 
Important to note, these regions are defined solely for a stationary gas. In an outflowing wind, the formation of molecules and dust also depends on the relative timescales of the outflow and the relevant reactions and the time-dependent diminishing concentration of condensible elements in the gas.

\begin{figure}
\vspace*{0.3cm}
\centering
\includegraphics[width=3.5in]{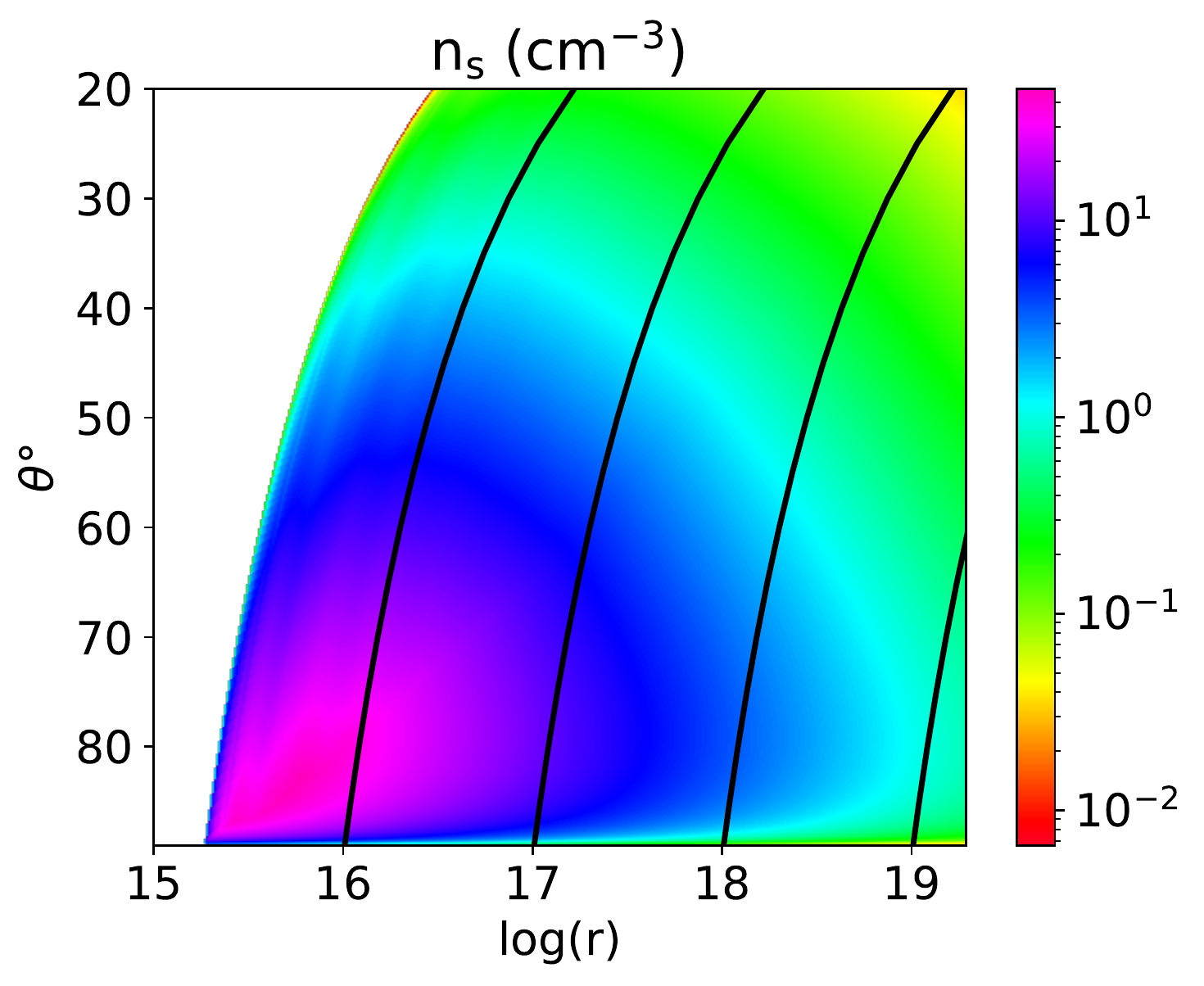}
\caption{\label{fig_nucleation}\footnotesize{The colormap shows the number density of dust precursors (seeds) in polar coordinates ($r, \theta$) coordinates, prior to condensation. The white region is free of seeds as the temperatures are too hot to sustain stable clusters along the course of the flow. The black lines on the figure represents the parabolic flow-lines with various starting/launch radius on the accretion disk, $r_0$.}} 
\end{figure}

\section{Kinetics of seed nucleation} 
\label{nucleation}

We assume the winds are launched dust-free from the surface of the accretion disk and follow the path given by Equation (\ref{motion}). Nucleation of silicates transpires through a set of chemical processes involving atoms and ions such as Si, O, Mg, Fe and SiO, H$_2$O, O$_2$ etc. \citep{gou12, sar13, gob16, bromley2016}. Other than the dimerization of SiO molecules, all processes in this chain are energetically favored, $i.e.$, the product is more stable than the reactants. This leads to the formation of stable dust precursors or seeds which in this case will have a form $\sim$ (Mg$_x$Fe$_{2-x}$SiO$_4$)$_2$ \citep{bromley2017}. We consider this as the seed-nucleus, which is the end-product of the nucleation process. The reaction rate of the bottle-neck for reaction for nucleation of (SiO-Mg-Fe) clusters is $\sim$ 10$^{-14}$ cm$^{3}$ s$^{-1}$ \citep{sar13} and for simplicity we take this as the general rate of nucleation, $\zeta \rm_{SiO-Mg}$ (= $\zeta$). 

The rate coefficient $\zeta$ is a function of the metal density, $n_z(r,\theta)$ = $nf_z$, where $f_z(r,\theta)$ is the fractional abundance of any metal (which is a constituent of the seed nuclei) at a given $r,\theta$. In case of silicate nucleation, we take it as the abundance of Si atoms in the gas, as the efficiency of the seed formation is limited by the availability of Si atoms. Let $f_{z0}$ be the fractional abundance at solar composition, $i.e.$, of the dust free gas. We calculate the number of seed nuclei in a co-moving unit-volume, $Y_\mathrm{s}$, (unit volume at $r_0$) and the seed number density, $n_\mathrm{s}$ in cm$^{-3}$, at a given coordinate $r_1,\theta_1$. 


\begin{equation}
\label{yield_nuc}
\begin{split}
& Y_{\mathrm{s}}(r_1,\theta_1) =  \int \mathrm{d}Y_{\mathrm{s}} =  \int^{\theta_1}_{\frac{\pi}{2}-\delta} \zeta \ f^2_{\rm{z}}(r, \theta) \ n^2(r, \theta) dt \\
& f_{\rm{z}} (r,\theta) = f_{z0} - \frac{n_s(r,\theta)}{n(r,\theta)}\\
& Y_{\mathrm{s}}(r_1,\theta_1) =  - \int^{\theta_1}_{\frac{\pi}{2}-\delta} \frac{ \zeta \ C_0^2 \ f^2_{\rm{z}} (r,\theta)}{\beta (r_0 GM_\mathrm{BH})^{1/2}} \frac{sin^{3/2}\theta}{cos^{1/2}\theta} e^{5(\theta-\pi/2)} \mathrm{d}\theta \\
& n_{\mathrm{s}}(r_1,\theta_1) = \int \frac{n(r_1,\theta_1)}{n(r,\theta)} \ \ \mathrm{d}Y_{\mathrm{s}}(r,\theta)
\end{split}
\end{equation}
Here we have used the `$\mathrm{d}t/\mathrm{d}\theta$' relation from Equation (\ref{eq_time}) and the gas density from Equation (\ref{eq_density_easy}). 
In the lower limit of the integral, the infinitesimal angular width, $\delta$ ($\sim$ $\Delta\theta$) represents the scale width over the surface of the accretion disk from which the outflow starts. In our study, we assumed a scale width of $h/R \sim$ 10$^{-2}$, which translates to an approximate inclination angle of $\sim$ 89.3$^\circ$ where the outflow launches. 

\begin{figure}
\vspace*{0.3cm}
\centering
\includegraphics[width=3.5in]{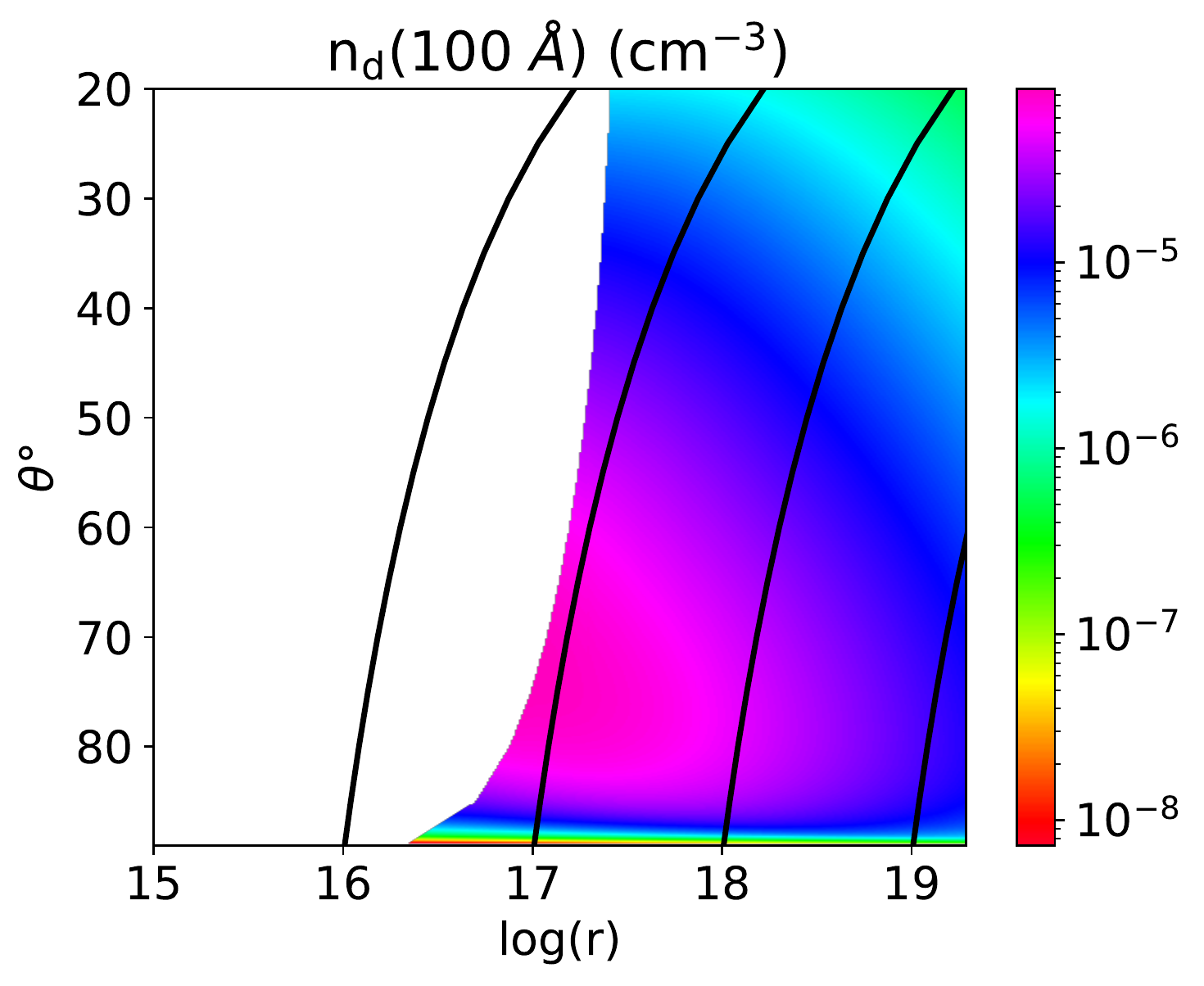}
\caption{\label{fig_coagulation}\footnotesize{The colormap shows the number density of 100~\AA\ silicate grains in polar coordinates ($r, \theta$) coordinates. The white region is free of dust as it is the boundary of the sublimation region. The lines in black on the figure represents the parabolic flow-lines with various starting/launch radius on the accretion disk, $r_0$.}} 
\end{figure}

The number density ($n_\mathrm{s}$) and the number of seeds in the co-moving volume ($Y_\mathrm{s}$) are related by the dilution in density along the line of the flow. In other words, $Y_\mathrm{s}$ gives the total number of seed-nuclei formed at a given $r_1,\theta_1$, from all the metals that were present in a unit volume of gas, when the flow started ($r_0$, $\theta \sim \pi/2$), whereas $n_\mathrm{s}$ calculates the number density of these seeds at $r_1,\theta_1$. 

Figure \ref{fig_nucleation} shows the the number density ($n_\mathrm{s}$) of the seed-nuclei in polar coordinates. The colormap clearly shows a high density of these dust precursors (seeds) near the large inclination-planes ranging between $\theta$$\sim$70$^\circ$ to $\theta$$\sim$85$^\circ$. As the winds move further outwards, there is a rapid decline in the number densities of these seeds caused by the (a) movement of the gas out of the nucleation zone, so formation of seeds is no longer viable; (b) decrease in the number density of the gas, leading to the thinning in the concentration of seeds; and (c) depletion of a large fraction of the refractory elements into molecules and seeds, so the rate of formation drops due to the lack of available condensible elements.

The maximum density of the seeds is attained within the radial range of 10$^{16}$ and  10$^{17}$ cm. The drop in number density of the seeds towards the outer radius is mainly due to the decrease in gas density, which also reduces the rate of nucleation. The white region on the figure represents the lack of any stable molecules or seeds. The wind in this region never passes through the nucleation zone, and hence always remains free of seed-clusters.


\begin{figure*}
\vspace*{0.3cm}
\centering
\includegraphics[width=3.5in]{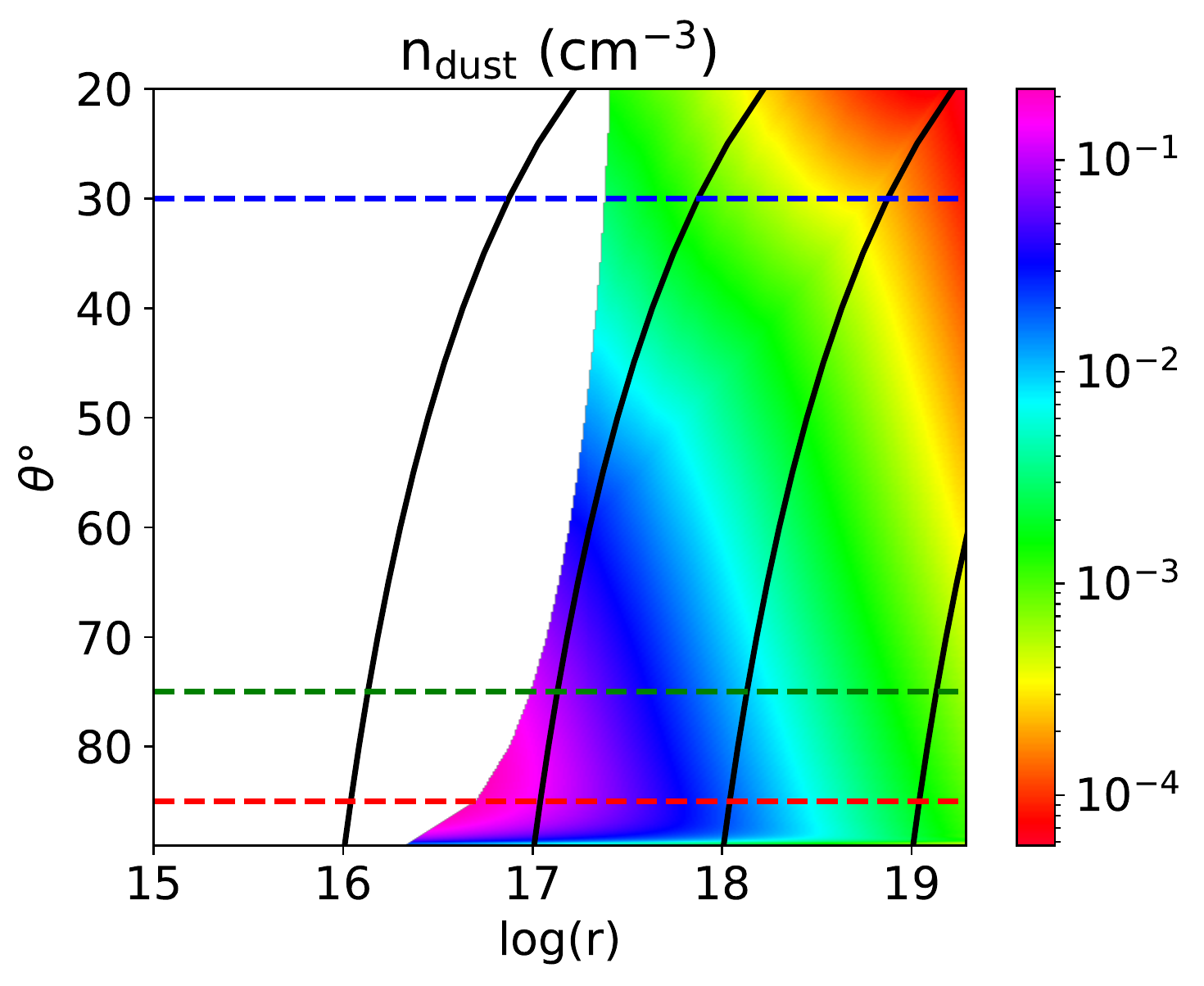}
\includegraphics[width=3.5in]{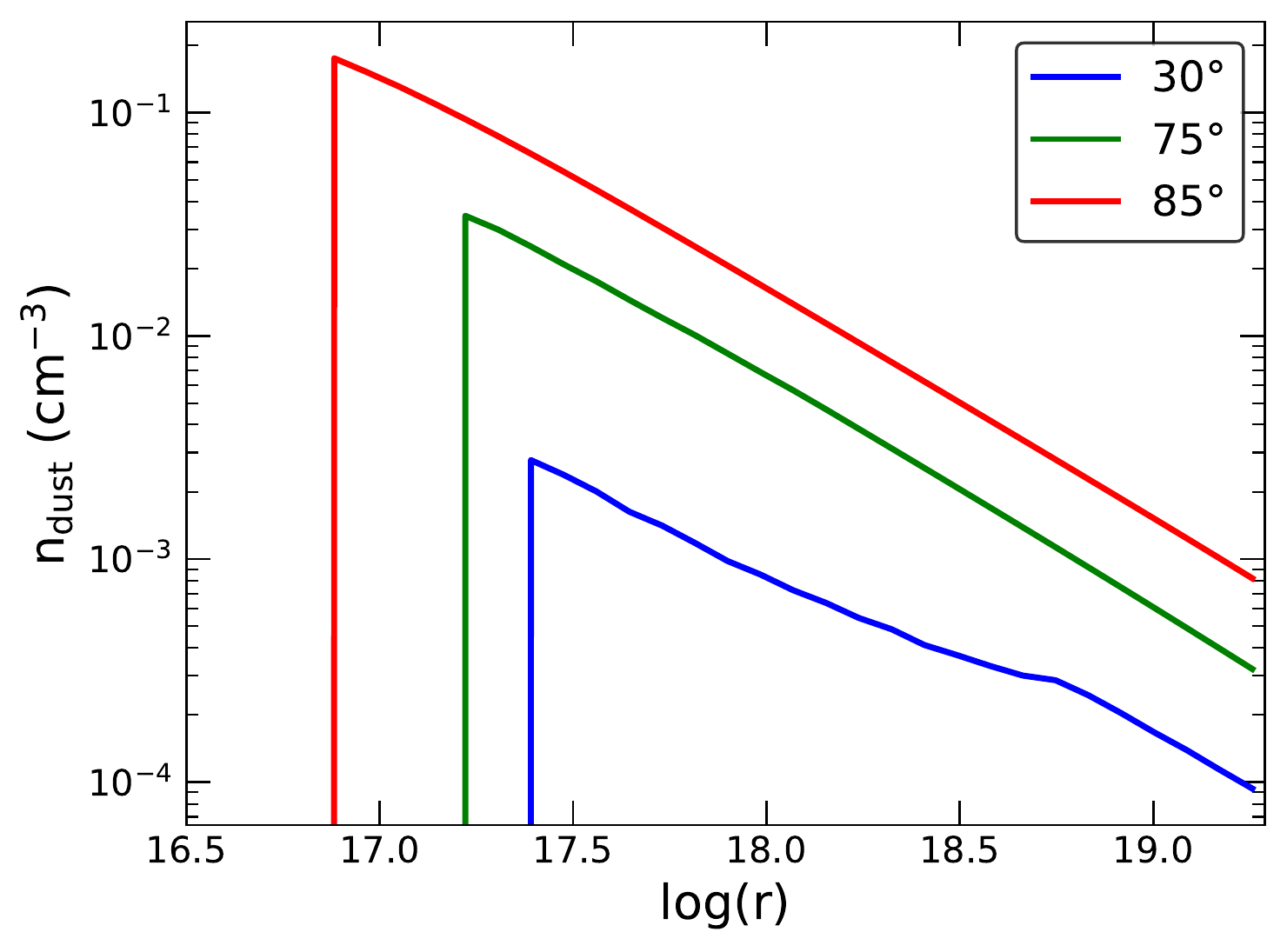}
\caption{\label{fig_dustdensity}\footnotesize{\textit{Left-panel:} The colormap shows the total number density of dust grains comprising of sizes ranging between 10~\AA\ and $\sim$ 200~\AA\ in polar ($r, \theta$) coordinates. The black lines on the figure represents the parabolic flow-lines with various starting/launch radius on the accretion disk, $r_0$. As seen in the figure, for each radius, the number density of grains increases at higher inclinations, $i.e.$, close to the plane of the accretion disk at $\theta \sim 90^\circ$. The figure shows that the dust forming region has a toroidal morphology. The three horizontal lines correspond to the inclination angles, for which the dust densities are shown in the figure on the right in the same color code. \textit{Right-panel:} The number density of dust grains is presented as a function of radial distance $r$ for inclination angles $\theta = 30^\circ, 75^\circ$ and $85^\circ$. The presence of a dust toroidal region at high inclination angles are further visible in the figure. }} 
\end{figure*}

\section{The number density of dust grains after seed coagulation}
\label{sec_dustdensity}

The number density and final size of dust grains are determined by the coagulation and accretion of the seeds. 

\begin{figure}
\vspace*{0.3cm}
\centering
\includegraphics[width=3.5in]{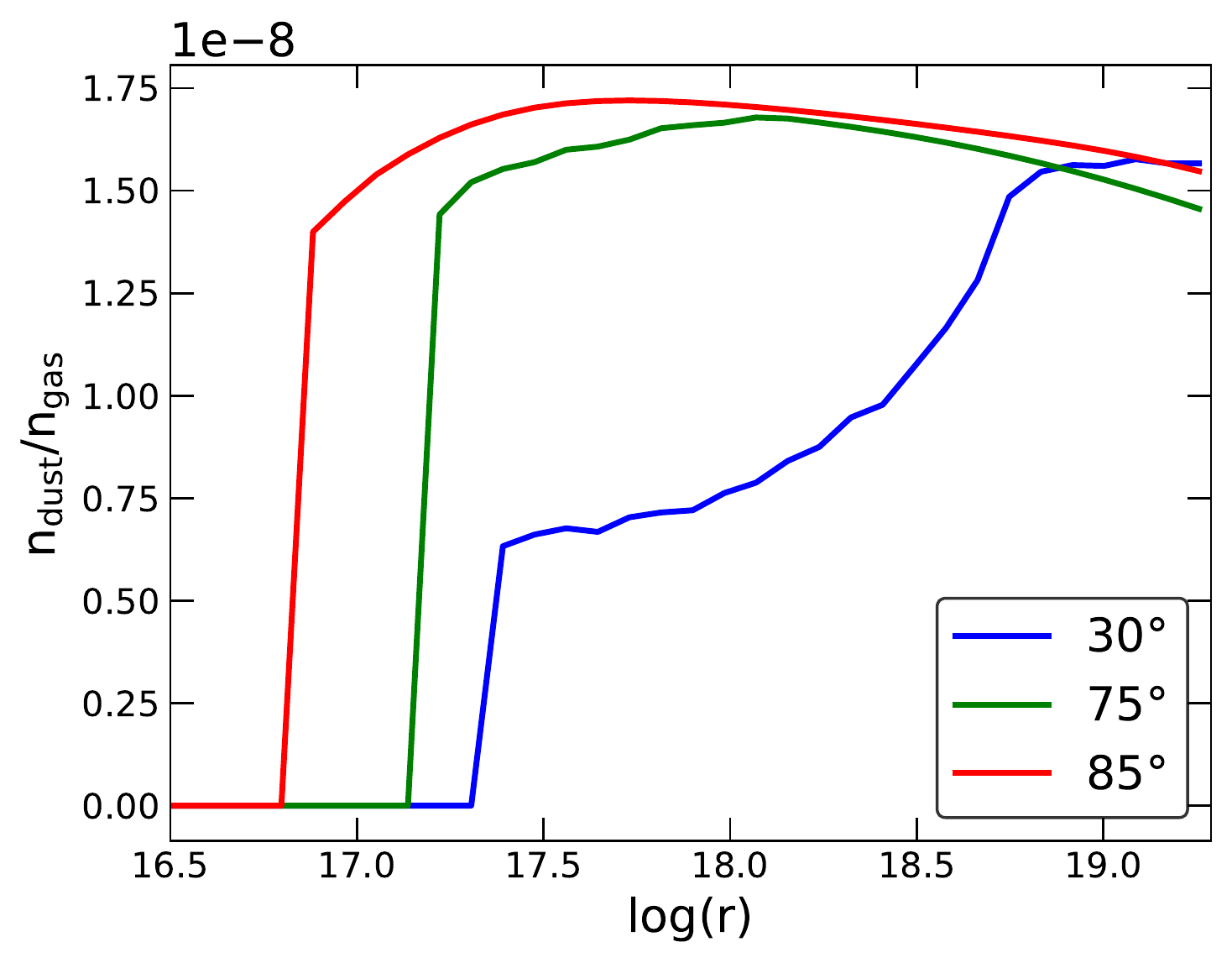}
\caption{\label{fig_ndustngas}\footnotesize{The dust to gas number density ratio along three different inclinations, $\theta = 30^\circ, 75^\circ$ and $85^\circ$, are presented as function of the radial distance from the central source, with reference to the gas number density shown in Figure \ref{fig_v_t_T_n} (bottom-left) and dust number density in Figure \ref{fig_dustdensity}.}} 
\end{figure}

Coagulation of seeds (dust precursors) and intermediate grains, and accretion of remaining metals on the grain surface are the two efficient routes of dust formation from molecules \citep{dwe11,sar15}. We consider the growth of seeds through coagulation using a volume conserved formalism described by \cite{mj05} and used by \cite{sar15}. The rate-coefficient for coagulation, $\xi_\mathrm{c}$ (cm$^3$ s$^{-1}$), between two particles of radii $a_i$ and $a_j$, in such environments is given by, 

\begin{equation}
\label{eq_coag}
\xi_\mathrm{c}(i,j) =  \pi (a_i + a_j)^2 \times \Big[\frac{8k}{\pi} \Big( \frac{T_i}{m_i} + \frac{T_j}{m_j}\Big)\Big]^{0.5} W_{ij}
\end{equation}
where, $T$ and $m$ are the kinetic temperatures and masses of the particles and $W_{ij}$ is the Van der Waal's force coefficient between them. $W_{ij}$ ranging between 2 to 10, we adopt an average value of 5 for this study \citep{mj05, sar15}. We consider the growth of seeds until they reach a size of 100 \AA, after which their growth proceeds through accretion.

For simplicity, we use an average coagulation rate as function of temperature. It corresponds to the coagulation between two 10 \AA\ silicate grains, with internal density of 3.2 g cm$^{-3}$, and express the rate coefficient as $\xi_\mathrm{c} \sim \eta \ T^{0.5}$. The dust temperature, depending on its size and the intensity of the incident radiation, can be kinetic or radiative in nature. For the coagulating grain, we have considered the kinetic temperature of the gas, given by Figure \ref{fig_v_t_T_n}, as the temperature of the coagulating seeds.

The first step is marked by the coagulation between two seed grains of size $a_0$, whose yields was calculated in Equation (\ref{yield_nuc}). The number of grains in a co-moving unit-volume, $Y_{\mathrm{d}}(a_1)$, and the number density, $n_{\mathrm{d}}(a_1)$ in cm$^{-3}$, of the resultant grain of size $a_1$ is given as,

\begin{equation}
\label{yield_dust}
\begin{split}
& Y_{\mathrm{d}}(a_1, r_1,\theta_1) =  \int \mathrm{d}Y_{\mathrm{d}} \\
& =  \int^{\theta_1}_{\frac{\pi}{2}-\delta} \eta \ T^{0.5}_\mathrm{g}(r,\theta) \ [Y_s^{\prime}(a_0,r, \theta)]^2 dt \\
& Y_s^{\prime}(a_0,r, \theta) = Y_{\mathrm{s}}(a_0,r, \theta) -  2 Y_{\mathrm{d}}(a_1,r,\theta) \\
& n_{\mathrm{d}}(a_1,r_1,\theta_1) = \int \frac{n(r_1,\theta_1)}{n(r,\theta)} \ \ \mathrm{d}Y_{\mathrm{d}}(a_1,r,\theta)
\end{split}
\end{equation}
where $Y_s^{\prime}(a_0)$ is the number of the remaining seeds of size $a_0$ in that co-moving volume of gas, at a given $r,\theta$, after the coagulation step. The factor 2 in the equation stands for the fact that each coagulation removes two seeds from the system. The relation between $n_\mathrm{d}$ and $Y_\mathrm{d}$, and the limits of the integration, follows the same analogy given in Section~\ref{nucleation}. 

Similar formalism was used to calculate the number density of grains of any size $a_\mathrm{i}$, until they reach 100 \AA\ in radius. For coagulation between grains of different sizes, the $Y_d^2(a_i)$ in Equation (\ref{yield_dust}) is replaced by $Y_d(a_i)Y_d(a_j)$. 

Figure \ref{fig_coagulation} shows the colormap for the distribution of the 100 \AA\ grains, in polar coordinates. The number density of dust grains in a given parcel of gas at ($r,\theta$), depends on the number density of the seeds present in the gas and on the timescales of coagulation compared to the time of the flow through a small volume around ($r,\theta$). Owing to the high density of seeds and the slower flow velocities near the large inclination angles, the density of 100 \AA\ grains are higher between $\theta$ $\sim$ 60$^\circ$ and 80$^\circ$.

The distribution of the total dust density, $n_\mathrm{dust}$ = $\sum n_\mathrm{d}(a_\mathrm{i})$, is shown in Figure~\ref{fig_dustdensity} where grains range between 10 and 100 \AA. The left-panel shows the colormap, where the dust number densities are found to be higher near the large-$\theta$ planes as expected. Importantly, the density distribution represents the combined result of simultaneous nucleation and coagulation processes, so dust density is higher where there is a higher concentration of seeds and slower rate of coagulation, while density is lower in the opposite case. The density of the dust grains declines rapidly with radius which is primarily due to efficient coagulation resulting from slower flow velocities towards outer radii. The white region represents the sublimation zone, which is free of dust. 

Only coagulation, however does not alter the dust to gas mass ratio, it only changes the size distribution of the grains. The accretion of the remaining metals on the surface of these grains, discussed in the following section alters the mass ratio of gas and dust. 

Figure~\ref{fig_dustdensity} (right panel) shows dust densities as a function of radial distance for three inclination angles. Both the figures confirm the presence of a dense toroidal region surrounding the accretion disk, with a scale height ranging up to $\theta$ $\sim$ 70$^\circ$. We identify this region with the observed dusty `tori' surrounding the AGN. 

The ratio of number densities of dust to gas, at three different viewing angles are presented in Figure \ref{fig_ndustngas} as a function of the distance from the central source. Close to the central BH lies the sublimation zone, where the ratio is zero due to the absence of dust. In the condensation zone, the dust to gas number density ratio is of the order of 10$^{-8}$. Along any inclination shown on the figure, the ratio increases fast along radial direction, and then gradually declines. The lowering of the dust to gas number ratio at larger radii indicates the lower efficiency of seed formation, owing to low densities. The ratio also reduces as the gas moves from high to low inclination planes (high to low $\theta$). This is mainly attributed to the condensation of small grains over time, into larger grains, thereby reducing the dust to gas number density ratio.



\begin{figure*}
\vspace*{0.3cm}
\centering
\includegraphics[width=3.5in]{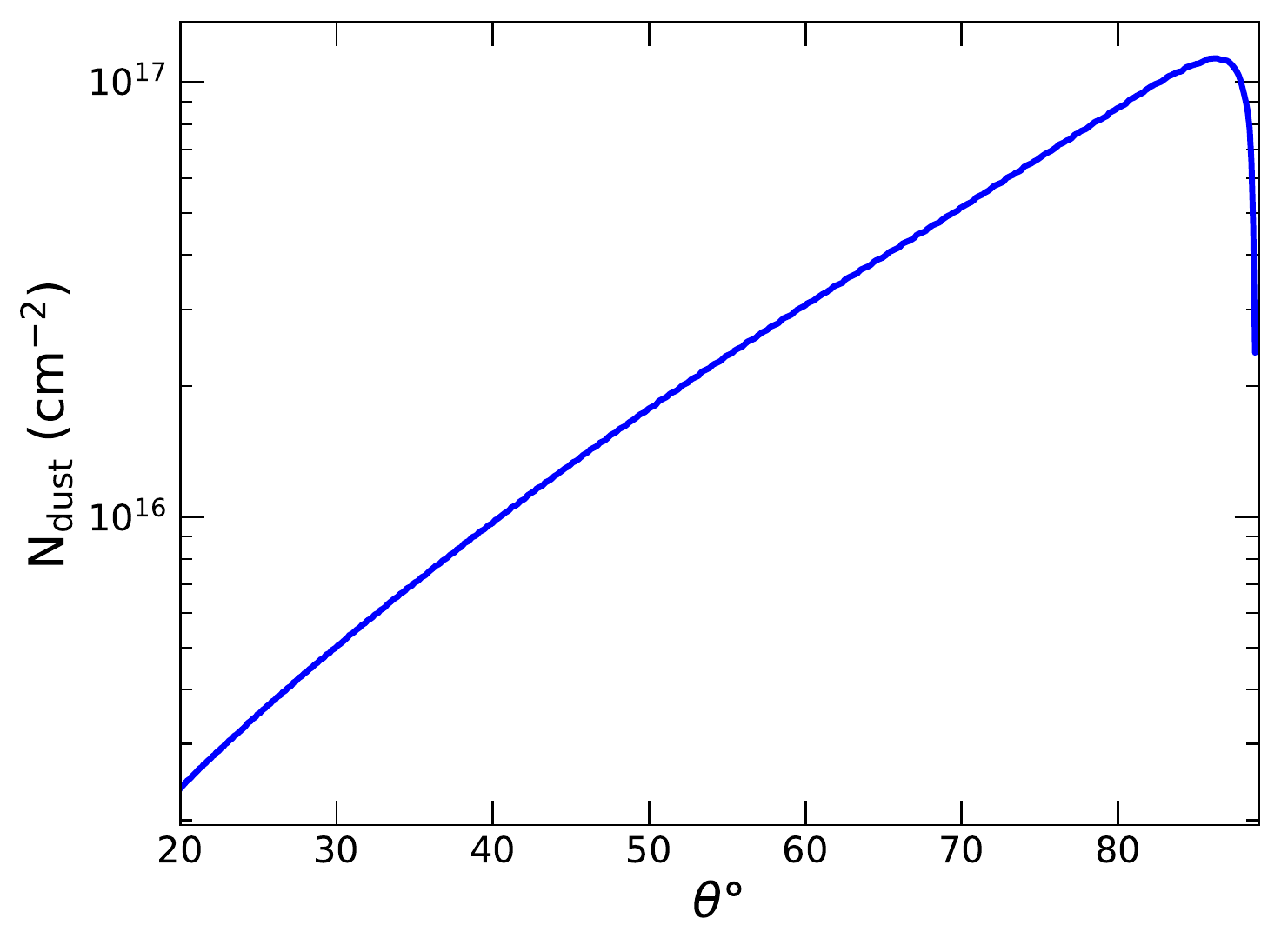}
\includegraphics[width=3.5in]{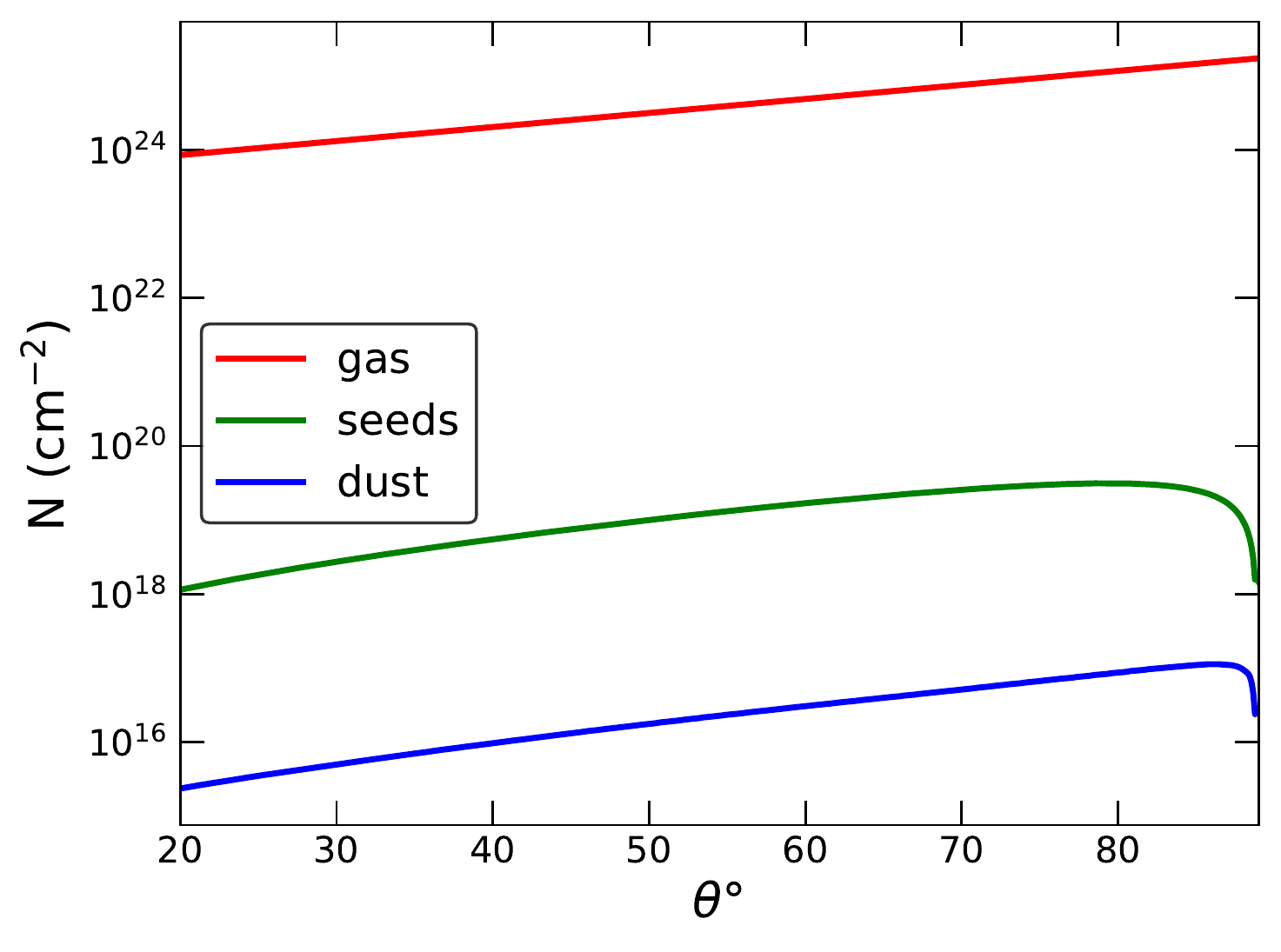}
\caption{\label{fig_columndensity}\footnotesize{\textit{Left-panel:} The figure shows the column density of dust grains as function of the inclination angle $\theta$. The dust column densities are larger at high inclinations, as seen in the figure. The steep drop in column densities towards low inclinations (low $\theta$, high-z) is attributed to the lowering of the gas densities and the scarcity of seeds owing to the coagulation of multiple seeds into one grain. \textit{Right-panel:} The column densities of the gas, the dust precursors (seeds) and the dust grains are presented as a function of the inclination angle $\theta$. }} 
\end{figure*}

Taking the central source as the origin, the column density ($N_\mathrm{dust}$) of dust grains along various lines of sight is given by, 

\begin{equation}
\label{eq_numbercolumn}
N_\mathrm{dust}(\theta) = \int n_\mathrm{dust}(r,\theta) \mathrm{d}r
\end{equation}
Figure~\ref{fig_columndensity} (left-panel) shows the dust column density as a function of inclination angles. The column densities also reflect the shape of the toroidal dusty zone, as the viewing angles along the equatorial directions (high-$\theta$, small-z) have a very high column density of dust which drops steeply with the decrease in inclination, $i.e.$, when viewed along the polar directions. Figure~\ref{fig_columndensity} (right-panel) shows the column densities of the gas, dust-precursors or seeds and the dust grains, as the function of inclination angles. All the processes related to dust formation, $i.e.$, nucleation of molecules and coagulation of grains, are most efficient when the wind is still at high inclination planes (zone C, Figure \ref{fig_zones}). In this low-inclination regions, the rate of coagulation drops due to the lack of available seeds. Moreover, the density of coagulating grains also declines because multiple seeds coagulate to form single grains. Therefore, the dust density in the wind is found to simply follow the behavior of gas density. 

\section{Size distribution following accretion}
\label{sec_grainsizes}


Once formed, the grains grow in size via accretion of the remaining metals on their surface. The efficiency of accretion is higher for large grain sizes as the grain surface area is larger and the sticking coefficient is close to unity. We consider the smallest size of the accreting grains as 100 \AA, which is the terminal size considered in the coagulation process.  

A parcel of gas at a certain coordinate ($r,\theta$) comprises of grains of various sizes. The distribution of grain sizes in that parcel depend on the following initial conditions: (i) physical conditions of the gas (ii) relative abundance of free metals compared to the accreting grains (iii) coordinates/epoch of formation of a certain grain along the path of the flow. 

The mass density of dust grains of radius $a$ is related to the number density of the grains by $\rho_{\rm{d}}(a)$ = $m_{\rm{gr}}(a) n_{\rm{d}}(a)$, where $m_{\rm{gr}}(a)$ is the mass of the individual grains of radius $a$. 
The rate of increase in mass density ($\rho_\mathrm{d}$) of grains of radius $a$, located at ($r,\theta$), due to accretion \citep{dwe11}, and the timescale of accretion, $\tau_{\rm{acc}} (r, \theta)$, is given by, 

\begin{equation}
\label{eq_drho}
\begin{split}
\frac{\mathrm{d\rho_{\rm{d}}}}{dt} & = \alpha_\mathrm{s} \ \pi a^2 \ \mu_z f_z \ n \ n_{\rm{d}} \Big(\frac{8 k T_\mathrm{gas}}{\pi \mu_z }\Big)^{\frac{1}{2}} \\
& \tau_{\rm{acc}}^{-1} = \frac{1}{\rho_{\rm{d}}} \frac{\mathrm{d}\rho_{\rm{d}}}{dt} \\
\end{split}
\end{equation}
where $\alpha_\mathrm{s}$ is the sticking coefficient, $n$ and $n_\mathrm{d}$ is the number density of the gas and the dust grains, $T_\mathrm{g}$ is the gas temperature, $\mu_z$ is the mean molecular weight of the accreting metals. The fraction of refractory elements remaining in the gas phase, $f_z (r,\theta)$, is given by Equation~\ref{yield_nuc}.
\begin{figure}
\vspace*{0.3cm}
\centering
\includegraphics[width=3.5in]{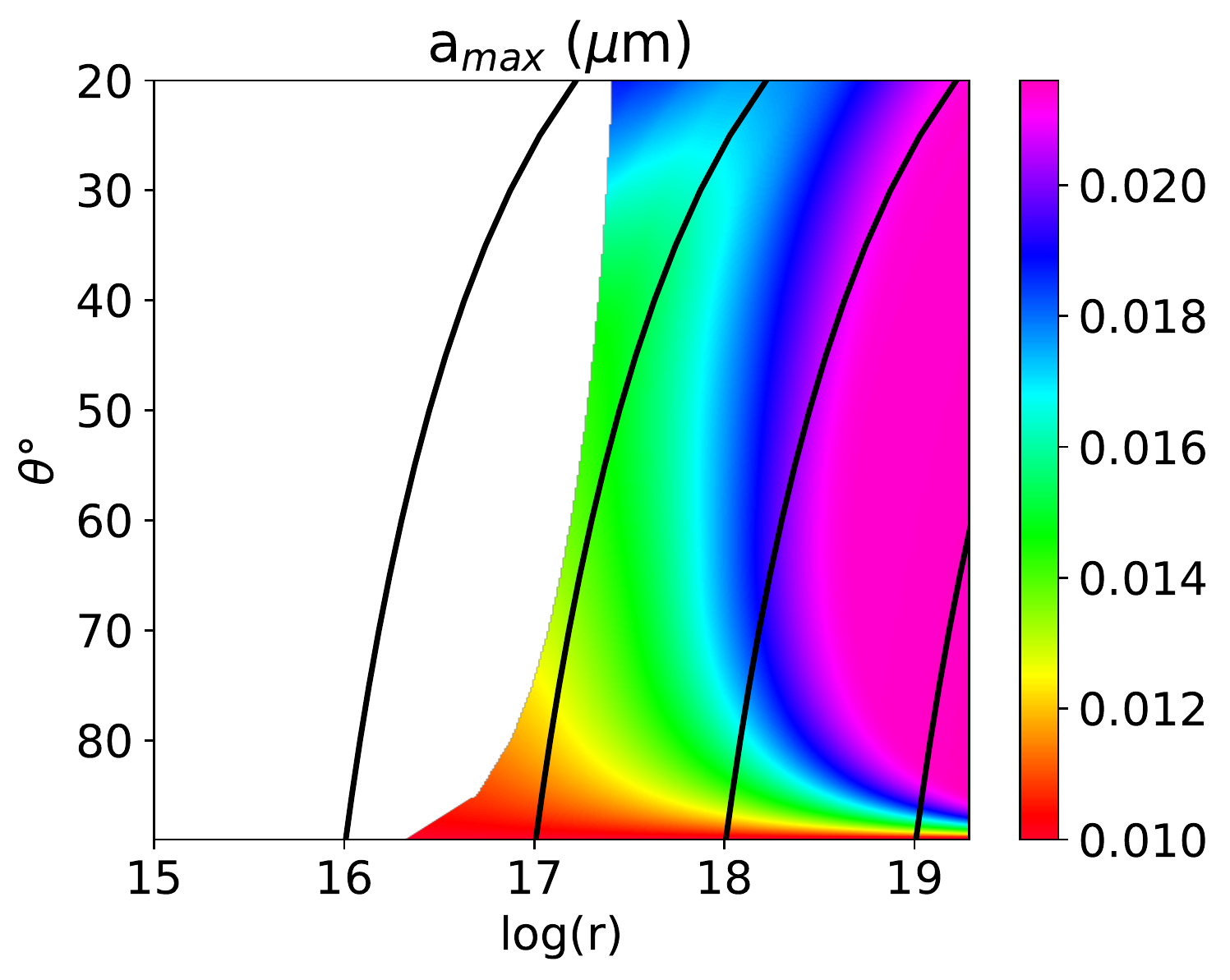}
\caption{\label{fig_grainsize}\footnotesize{The colormap shows the maximum grain sizes at a given region ($r, \theta$) where the grains have grown by accretion of free metals in the gas along the flow-lines. The lines in black on the figure represents the parabolic flow-lines with various starting/launch radius on the accretion disk, $r_0$.}} 
\end{figure}

The number density of the grains does not change via accretion and we simplify Equation (\ref{eq_drho}) as, 

\begin{equation}
\frac{\mathrm{d}a}{dt} = \frac{\alpha_\mathrm{s} f_z n}{\rho_\mathrm{gr}} \Big(\frac{k \mu_z}{2 \pi}\Big)^{\frac{1}{2}} [T_g(r,\theta)]^{\frac{1}{2}}
\end{equation}
where $\rho_\mathrm{gr}$ is the density of each grain in g cm$^{-3}$ (3.2 g cm$^{-3}$ for silicates). 

However, the remaining metals in an unit volume of gas are accreted uniformly among all the dust grains present in that volume. So the rate of growth of a single grain is corrected by the grain density as,  
\begin{equation}
\frac{\mathrm{d}a}{dt} = \frac{\alpha_\mathrm{s} f_z n}{\rho_\mathrm{gr} ||n_d||} \Big(\frac{k \mu_z}{2 \pi}\Big)^{\frac{1}{2}} [T_g(r,\theta)]^{\frac{1}{2}}
\end{equation}
where $||n_d||$ is the dimensionless quantity representing the grain density.  

Using the values of $n$ and $\mathrm{d}t$, from Equation (\ref{eq_density_easy}) and \ref{eq_time}, the final radius of a grain of initial size $a_c$ is, 
\begin{equation}
\begin{split}
 & a(r_1,\theta_1) = \int^{a}_{a_c} da = \\
& - \int^{\theta_1}_{\frac{\pi}{2}-\delta} \frac{\alpha_\mathrm{s} C_0 f_z(r,\theta)}{\rho_\mathrm{gr} \beta ||n_d||} \Big(\frac{k \mu_z r_0}{2 \pi G M_\mathrm{BH}}\Big)^{\frac{1}{2}}  \frac{sin^{3/2}\theta}{(1-cos\theta) cos^{1/2}\theta} \\
&   \ \ \ \ \ \ \ \ \ \ \ \ \ \ \ \ \ \ \ \ \ \ \ \ \ \ \ \ \ \ \ \ \ \ \ [T_g(r,\theta)]^{\frac{1}{2}} e^{\frac{5}{2}(\theta-\pi/2)} \mathrm{d}\theta \\
\end{split}
\end{equation}
where $a_c$ = 100~\AA. 

The largest radii of the dust grains, $a_{max}$($r,\theta$) in presented in Figure \ref{fig_grainsize}, resulting from accretion of the remaining metals. The largest grains are found to be around $\sim$ 220 \AA\ in radius, which is present in the outer regions with radius larger than 10$^{18}$ cm. The figure indicates in general that the grains are larger towards the longer radii, outwards of 10$^{18}$ cm. Inefficient nucleation in this region leads to a larger abundance of residual metals in the gas. This in turn make accretion more efficient. Slower flow-velocities (Figure \ref{fig_v_t_T_n} (top-left)) further aid the grain growth in these regions. However, grain sizes are limited to the amount of metals available for accretion. Hence, as the wind moves further away from the accretion disk (along the flow), most of the metals are already locked up into dust and the accretion rate slows down significantly. Overall, the newly formed dust grains range in radii between 10 to 220 \AA. 


\begin{figure}
\vspace*{0.3cm}
\centering
\includegraphics[width=3.5in]{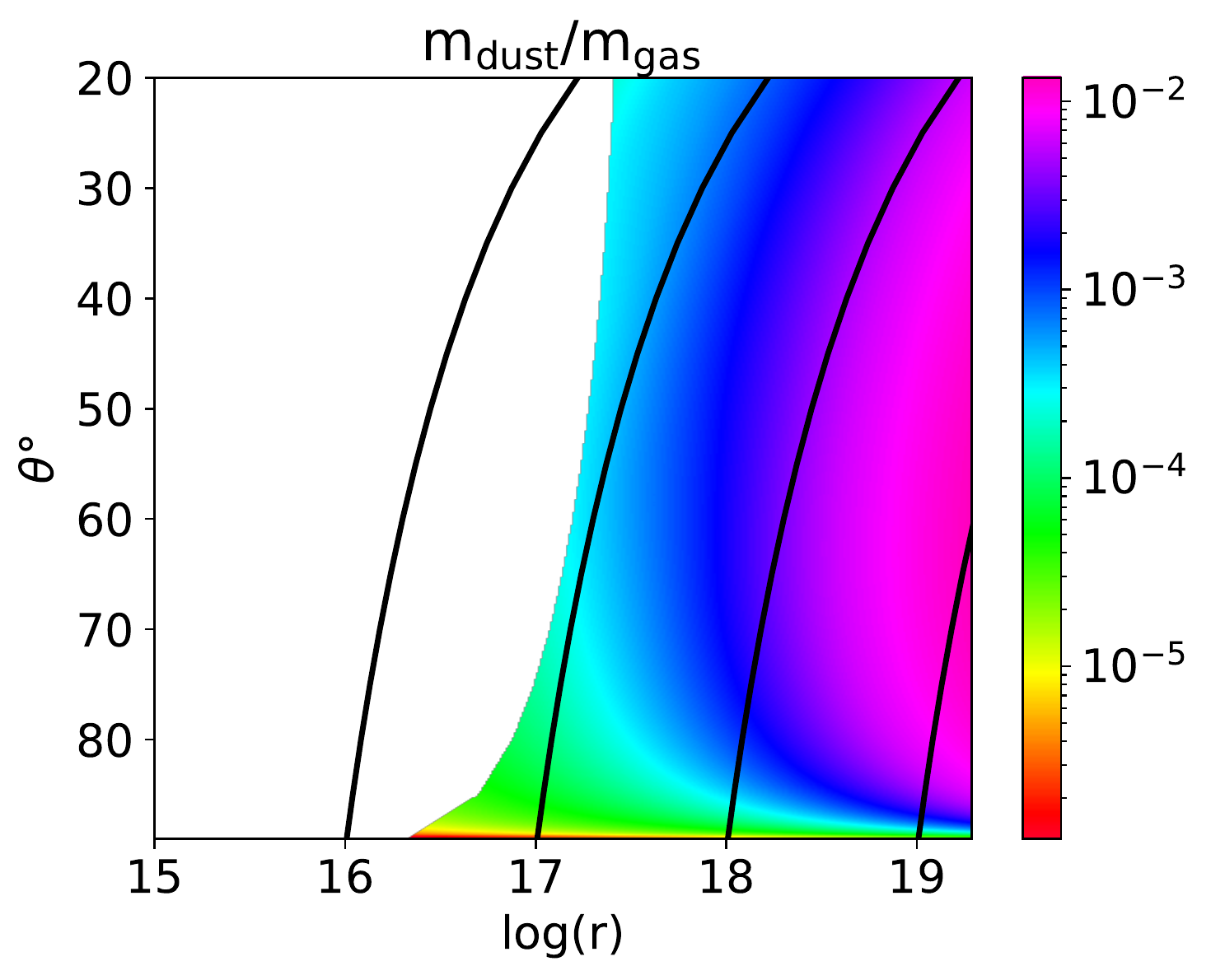}
\caption{\label{fig_dusttogas}\footnotesize{The colormap shows the dust to gas mass ratio in polar coordinates ($r, \theta$). The white region is the sublimation zone, and the black lines indicates flow lines with various launch radius $r_0$. The mass ratio seems to be larger towards the longer radii, outwards of 3$\times$10$^{18}$ cm, where the ratio is $\sim$ 10$^{-2}$. }} 
\end{figure}

Summing over grains of all sizes present in an unit volume of gas at a given ($r,\theta$), the mass density of the dust grains is calculated. Figure \ref{fig_dusttogas} shows the ratio of dust to gas mass densities in polar coordinates. The ratio is shown to vary between 10$^{-5}$ to 10$^{-2}$ where the maximum attained around $r$ $\sim$ 6$\times$10$^{18}$ cm and $\theta \sim$ 60$^\circ$. Moreover, the dust to gas mass ratio along the 2D plane follows a similar distribution as that of $a_{max}$ (Figure \ref{fig_grainsize}), thereby verifying a direct connection between the efficiency of accretion and the mass density of dust grains. 

\begin{figure*}
\vspace*{0.3cm}
\centering
\includegraphics[width=3.5in]{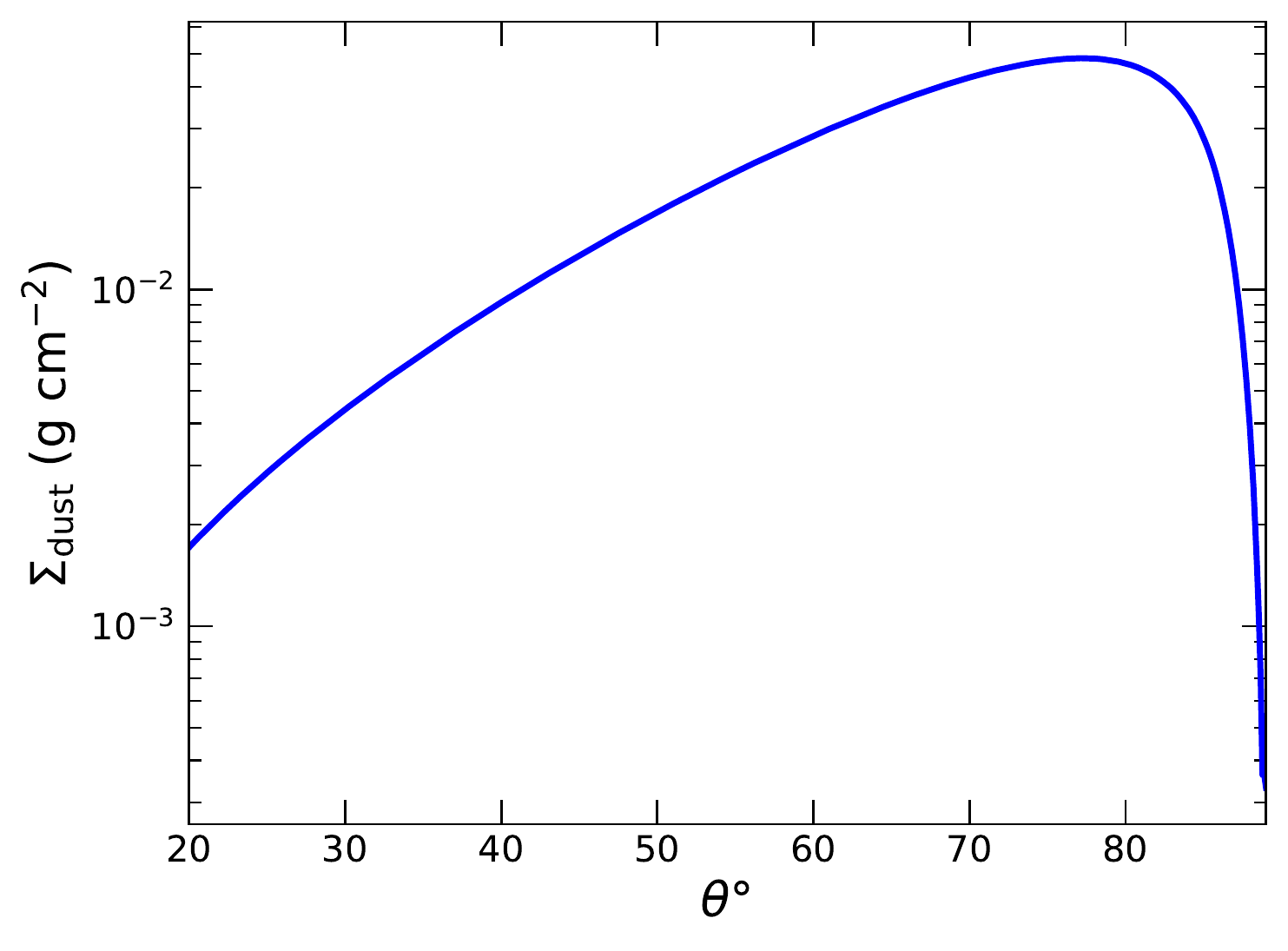}
\includegraphics[width=3.5in]{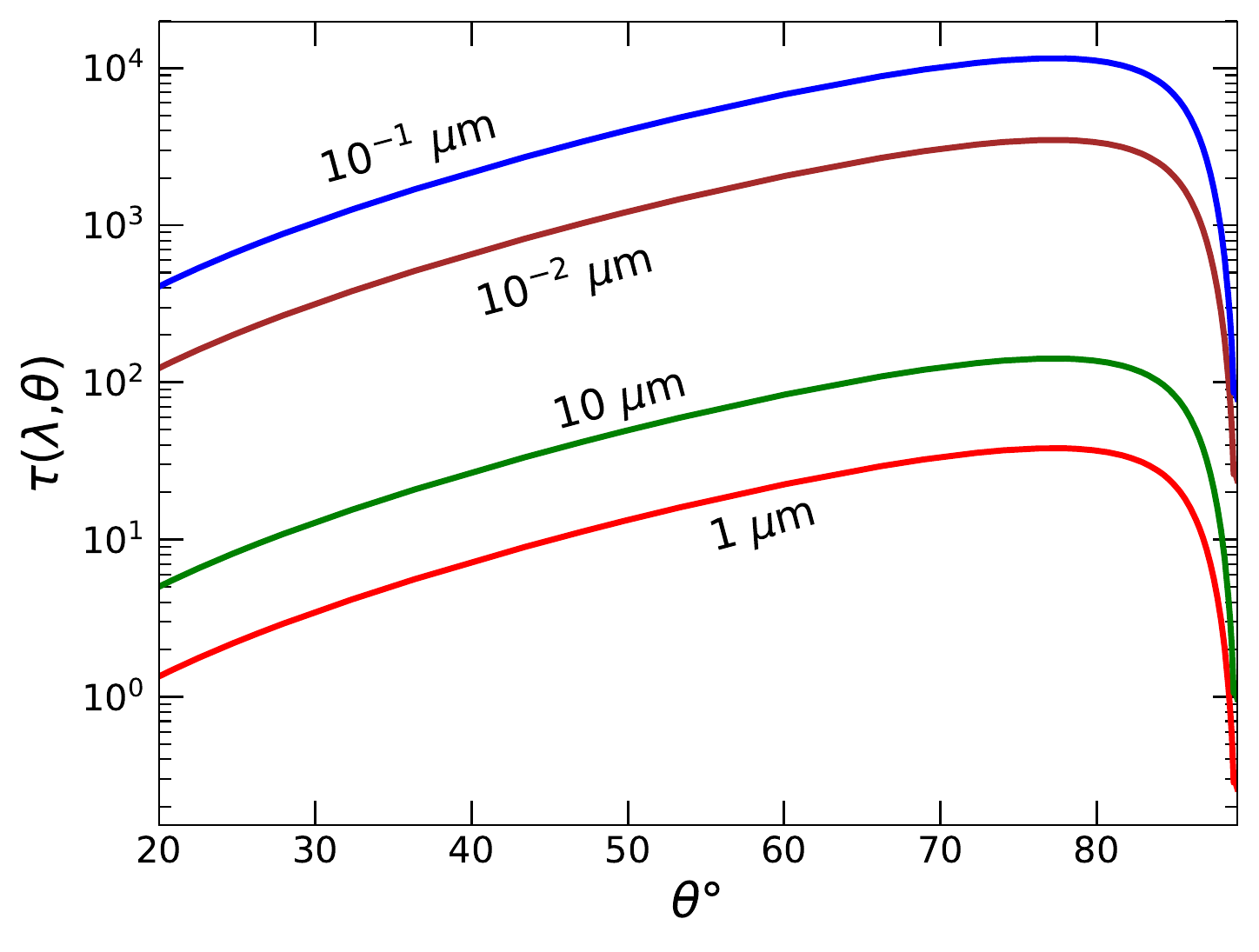}
\caption{\label{fig_masscolumn}\footnotesize{\textit{Left-panel:} The mass column density of the dust grains is shown as a function of inclination angle $\theta$. The column density of dust masses increases rapidly from the time of launch at $\sim$ $90^\circ$ and reaches its peak around $\sim$ $80^\circ$. Following that, as the gas flows towards low inclinations (low $\theta$, high-z) the number density of dust declines and mass column density also likewise. This further confirms that new dust formation in these regions are negligible. \textit{Right-panel:} The optical depth $\tau (\lambda)$ of the medium, resulting from the dust formation, is presented as a function of the inclination angle $\theta$, for four wavelengths, $\lambda$ = 0.01, 0.1, 1 and 10 $\mu$m.  }}
\end{figure*}


\begin{figure}
\vspace*{0.3cm}
\centering
\includegraphics[width=3.5in]{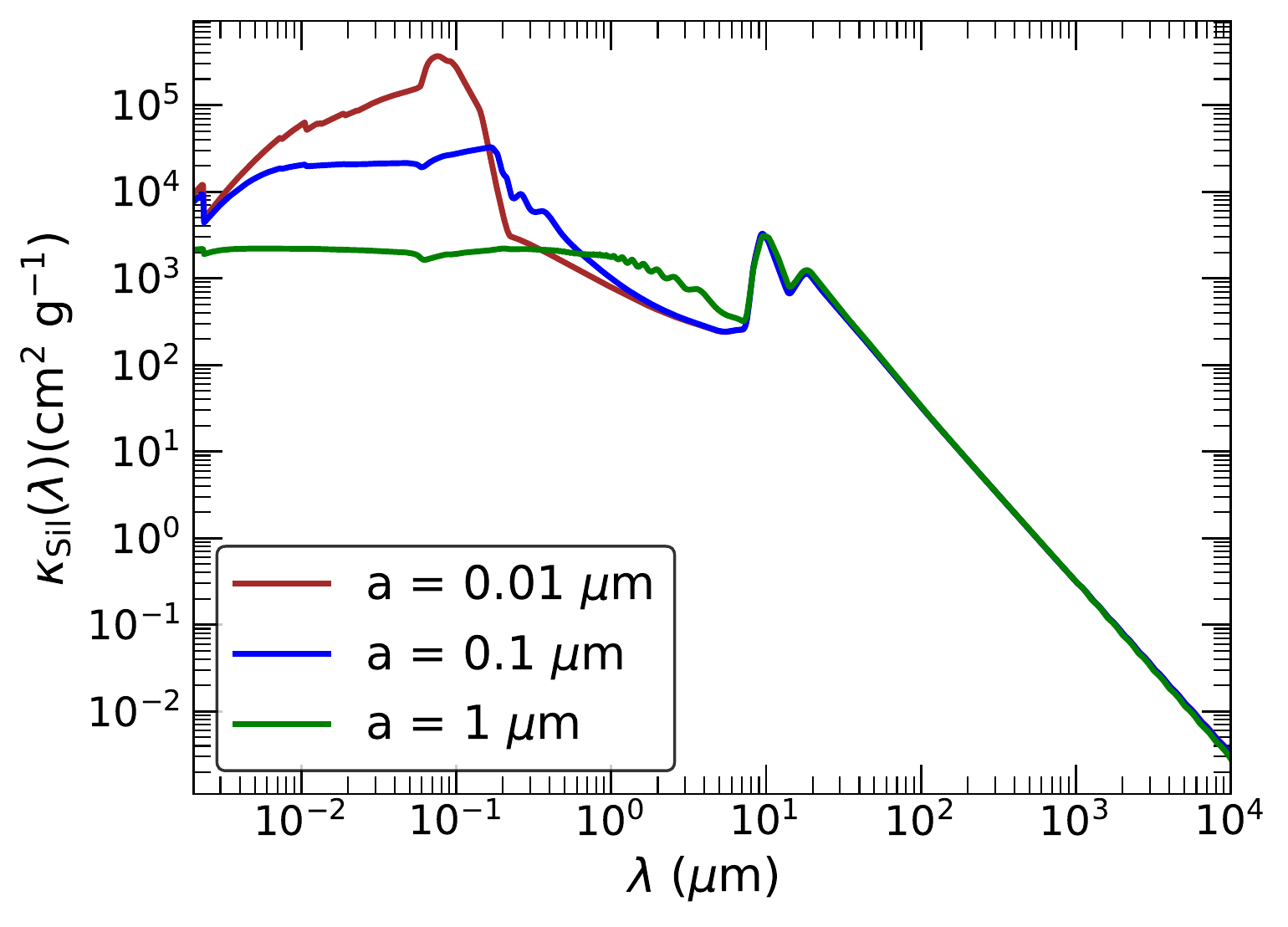}
\caption{\label{fig_ksil}\footnotesize{The mass absorption coefficient of astronomical silicates, derived from \cite{wei01}, is shown for three different grain radii, $a$ = 0.01, 0.1 and 1 $\mu$m, which is the range in which majority of the grains are present. }} 
\end{figure}

\section{The dust masses and the optical depths}
\label{sec_dustmass}
In this section we calculate the dust masses at any given epoch, the rate of dust formation and the optical depths at various wavelengths.

Assuming the winds to be in steady state, the total amount of dust present at any given epoch is just the integrated sum over all grain sizes.

Taking the spherical volume encompassed by the radius of the black hole influence and assuming azimuthal symmetry, the total mass of dust at any given epoch is calculated as,
\begin{equation}
\label{totalmass}
M_\mathrm{tot} = 4 \pi \sum_i \iint n_\mathrm{d}(a_\mathrm{i},r,\theta) m_\mathrm{d}(a_\mathrm{i}) r^2 sin\theta \ \mathrm{d}r \ \mathrm{d}\theta
\end{equation}
where $m_\mathrm{d}(a_\mathrm{i})$ is the mass of a dust grain of radius $a_\mathrm{i}$. Integrating over the $r$ and $\theta$ coordinates, the total mass of dust present within the sphere of influence at a given epoch is found to be $\sim$ 1.5 $\times$ 10$^4$ \Ms\ for the values of the black hole mass and dimensionless accretion rates employed throughout ($M_{\rm BH} = 10^7$\Ms\ and $\dot m = 1$; see section \ref{sec_phmodel}). The column density in mass (g cm$^{-2}$), along any given inclination is calculated as,

\begin{equation}
\label{masscolumn}
\Sigma_\mathrm{dust}(\theta) = \sum_i \int n_\mathrm{d}(a_\mathrm{i},r,\theta) m_\mathrm{d}(a_\mathrm{i}) \mathrm{d}r
\end{equation}

Figure \ref{fig_masscolumn} (left-panel) shows the mass column densities as a silicate grains function of the inclination angle. The behavior of $\Sigma_\mathrm{dust}$ resembles that of the column densities ($N_\mathrm{dust}$) shown in Figure \ref{fig_columndensity}. The maximum of the $\Sigma_\mathrm{dust}$ is around $\sim$ 85$^\circ$ while for $\Sigma_\mathrm{dust}$ the same is around $\sim$ 80$^\circ$. This small variation is caused by the growth of grains through accretion as the winds flow out towards lower inclination planes.

The rate of dust formation ($\dot{M}_\mathrm{dust}$) is calculated by integrating the rate of formation of the different sized grains over the entire volume. Using Equations \ref{eq_time} and \ref{totalmass}, $\dot{M}_\mathrm{dust}$ is given by,

\begin{equation}
\label{ratedustmass}
\dot{M}_\mathrm{dust} = 4 \pi \sum_i \iint \frac{\mathrm{d}n_\mathrm{d}(a_\mathrm{i},r,\theta)}{\mathrm{d}t} m_\mathrm{d}(a_\mathrm{i}) r^2 sin\theta \mathrm{d}r \mathrm{d}\theta
\end{equation}

In this case, the dust formation rate is calculated to be $\sim$ 0.37 \Ms\ yr$^{-1}$ (for $M_{\rm BH} = 10^7$ \Ms\ and $\dot m = 1$). Assuming the lifetime of the AGN to be around 1 Gyr, therefore it will lead to the formation of $\sim$ 10$^8$ \Ms\ of dust. However, one has to bear in mind that over this timescale, both the mass of the black hole and the dimensionless accretion/wind rates may change considerably. Moreover, the wind can actually contain pre-existing dust which gets reprocessed in the winds, instead of new dust formation. While the above number could be an absolute upper limit, further insights on the evolution and the dynamics of the galaxy is necessary to determine how much dust will be formed in the AGN winds over the entire galactic lifetime.

Should the accreting gas contain already a certain amount of dust, the winds likely be reaching dust saturation quickly, $i.e.$, all available metals being aggregated onto dust grains, a situation with likely observable consequences. On the other hand, a lower abundance of available metals could reduce the rate of chemical processes considerably, which would therefore lead to inefficient formation of new dust.





For the specific conditions considered in producing the figures of dust distribution given above, we can compute the winds' optical depth, $\tau$, for different wavelengths $\lambda$ and viewing angle $\theta$. This is given by the expression,

\begin{equation}
\label{eq_tau}
\tau(\lambda, \theta) = \sum_i \kappa(a_\mathrm{i},\lambda) \int n_\mathrm{d}(a_\mathrm{i},r,\theta) m_\mathrm{d}(a_\mathrm{i}) \mathrm{d}r
\end{equation}
where $\kappa(a_\mathrm{i}\lambda)$ is the mass absorption coefficient (cm$^2$ g$^{-1}$) of astronomical silicates, derived from \cite{wei01}. Figure \ref{fig_masscolumn} (right-panel) shows the variation of $\tau(\lambda, \theta)$ for $\lambda$ = 0.01, 0.1, 1 and 10 $\mu$m, as a function of the inclination angle $\theta$. The absorption coefficients for astronomical silicates is presented in Figure \ref{fig_ksil}, to show the variation of $\kappa(a,\lambda)$ as functions of grain size and wavelength. Since the optical depth is the integral of $\kappa(a_\mathrm{i}\lambda) \times \rho_d(r,\theta)$ along the observer's line of sight ($\rho_d$ is the mass density), the wavelength dependence of the absorption coefficient plays significant role in shaping the optical depth. Figure \ref{fig_ksil} shows that for longer wavelengths ($\lambda > 2 \mu$m), the absorption coefficient is almost independent of the grain size, while in case of shorter wavelengths the resultant size distribution would make a difference. The mass absorption coefficient for silicates has local maxima at 9.7 and 18 $\mu$m, a feature reflected in the larger optical depth at 10 $\mu$m in Figure \ref{fig_masscolumn} (right-panel) and also in the AGN spectra.


The optical depth increases with $\theta$, which is expected, given the increase of the wind column at higher inclinations. More significant is the depth increase with decreasing wavelength $\lambda$, which makes these winds at $\lambda \lsim 1 \; \mu$m highly opaque even for $\theta \simeq 20^\circ$. The optical depth decreases almost monotonically for $\lambda > 1 \;\mu$m with exception of the region around the silicate transitions at $\lambda \simeq 9.7, 18 \;\mu$m, implying the possibility of discerning these features in emission or absorption depending on the observer inclination angle and hence with the classification of an AGN as Syefert 1 or 2. This notion fits well with the analysis of these features in both types of Seyferts by  \cite{shi2006}.

Important to note, most of the observed spectra from the AGNs are not spatially resolved. Therefore, the observed spectra are generally an average over the entire cross-section of the region that we have modeled in this study. Detailed radiative transfer calculations, taking into account the absorption and re-emission of secondary photons, are required to model spectra which can be compared to the observations. We will present such radiative transfer calculations in a future article.

\begin{figure*}
\vspace*{0.3cm}
\centering
\includegraphics[width=3.5in]{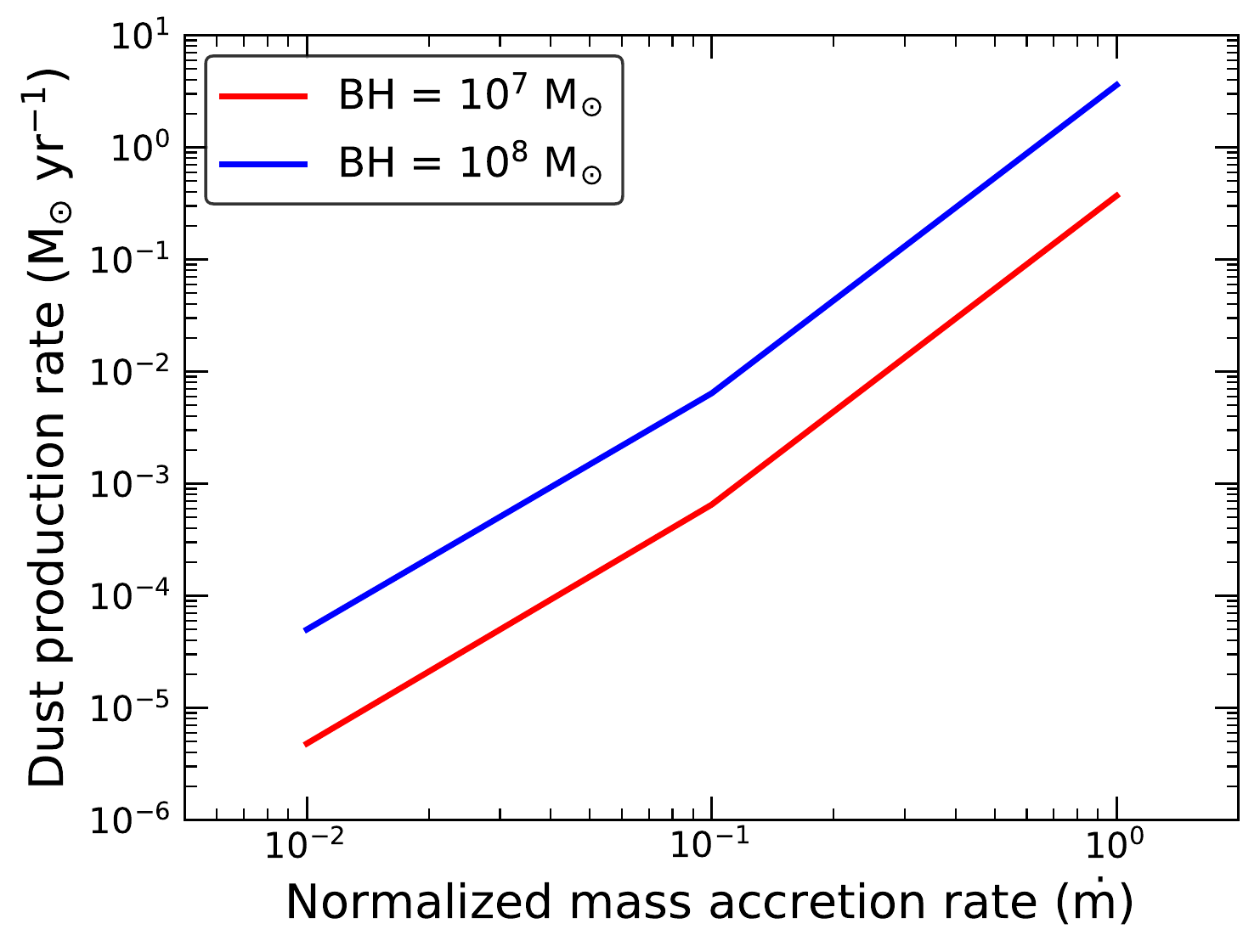}
\includegraphics[width=3.5in]{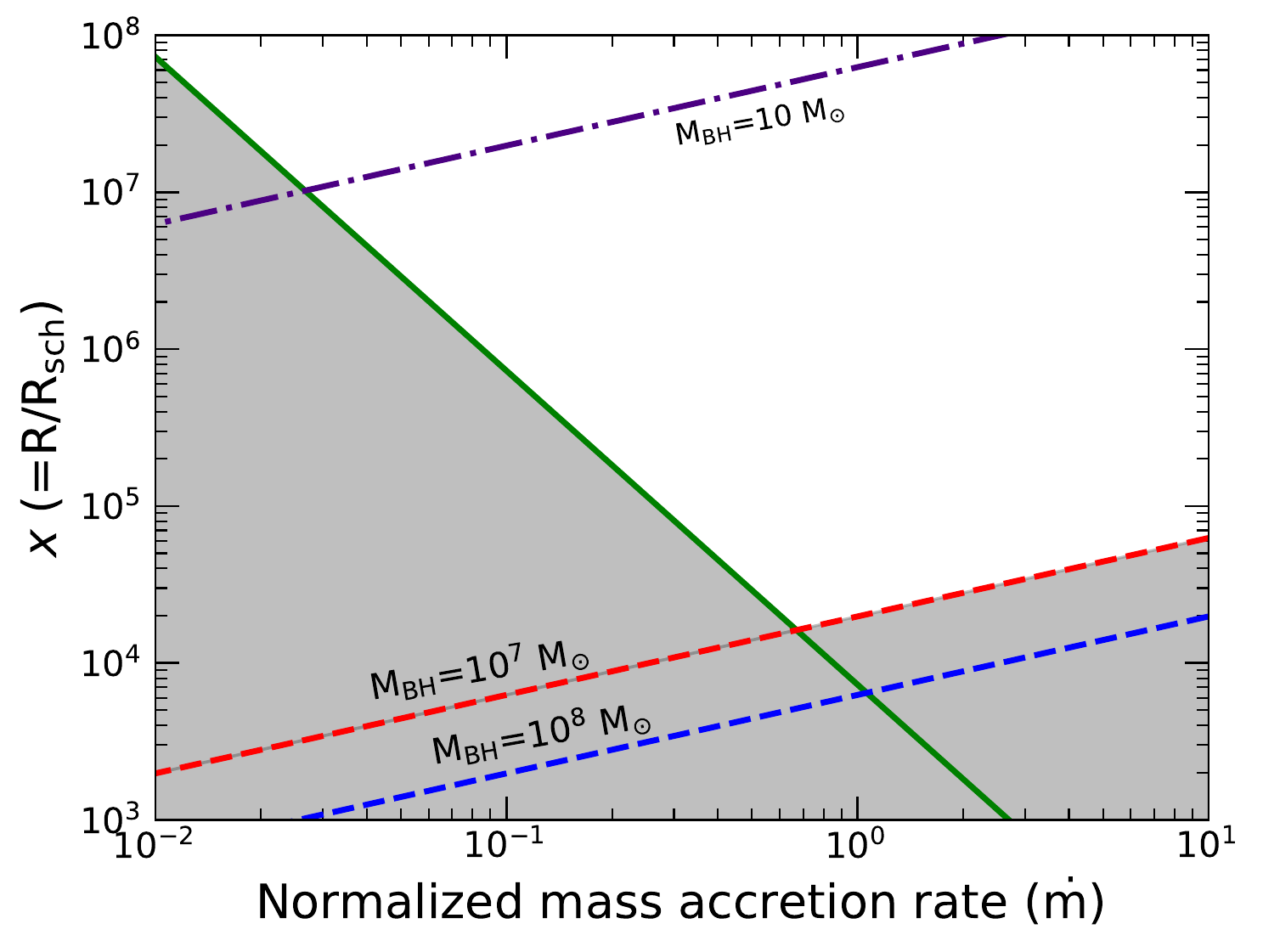}
\caption{\label{fig_parameterspace}\footnotesize{\textit{Left-panel:} The dust mass flux in solar masses per year, as a function of the (normalized) accretion rate, $\rm \dot{m}$, onto the BH (assumed equal to the wind mass flux at the smallest disk radii) for two different BH masses, $M = 10^7$ \Ms\ (red) and $10^8$ \Ms\ (blue). For $\dot m \simeq 1$ the dust mass flux is comparable to the accretion rate onto the BH (we remind the reader that the wind mass flux increases like $\simeq X^{1/2}$ with radius, so the local mass flux in the wind at $X \sim 10^5 - 10^6$ is much higher than that needed to power the BH emission). \textit{Right-panel:}  The figure shows the relation between the dimensionless length ($X$) and the normalized mass accretion rate $\dot m$} to identify the limiting conditions which are necessary for dust formation and survival. The green line depicts the line where the ratio of the timescales of chemical reactions and the flow time of the wind is equal to unity. The dotted lines represent the limiting conditions for dust survival, based on the sublimation temperature, for different BH masses. The region in white, therefore, represents the parameter space, where the values of $\dot m$ and $X$ together support the formation and survival of dust grains. }
\end{figure*}

\section{Dependance on accretion rate and blackhole mass}
\label{dependancies}

To examine the dependance of the results on the mass accretion rate and the BH mass, we have computed six cases in total, with mass accretion rates $\dot m$ = 0.01, 0.1 and 1, and BH  mass of  10$^7$ and 10$^8$ \Ms. The changes are manifested in the gas density and the luminosity of the source, while the spectral shape of the source is kept unaltered.

In Figure \ref{fig_parameterspace} (left-panel), we show the dust mass flux ($\dot M_{\rm{dust}}$) in solar masses per year as a function of $\dot m$ for these two different BH masses. For a given value of $\dot m$, the dust mass flux was found to scale proportional to the BH mass, just like the mass flux in the wind, in agreement with the qualitative arguments given above. However, its normalization has a stronger dependence on the dimensionless wind flux, which we estimated as $\dot m_{\rm dust} \propto \dot m^{b}$; where $b$ is between 2 and 2.5. One should note that the ordinate of this figure is given in solar mass per year rather than in dimensionless units; as such it is proportional to the BH mass, as the calculations indicate and the figure shows. In case of the dust mass production rate in dimensionless units, the figure would be independent of the BH mass.

The steep dependance on $\dot m$ is due to the fact that dust production requires the formation of seeds, whose number density is originally too small for the dust growth time scale to compete with the plasma flow time scale, as $\dot m $ decreases (remember that the ordinate in Figure is in absolute and not normalized values of $\dot m$). Moreover, the formation of dust grains depends of a series of interactions, the rate of which are proportional to the square of the density $n^2$, while $n$ is linearly proportional to $\dot m$. The relation between $\dot M_{\rm{dust}}$ and $\dot m$ in Figure \ref{fig_parameterspace} therefore is a curve of slope $\sim$ 2 in log-scales, which confirms the linear relation between gas density and $\dot m$.


Incidentally, near $\dot m \simeq 1$, the value of $\dot M_{\rm dust}$ can be (for the conditions considered) larger than the mass accretion rate that powers the BH luminosity. This may sound at first sight impossible, however, we would like to remind the reader that under the wind scalings assumed in our calculations, especially its radial dependence, $n(r) \propto 1/r$,  the wind mass flux has the following dependence $\dot M \propto r^2 V_K(r) n(r) \propto \dot m M X^{1/2}$ ($X$ =$R/R\rm_{sch}$), i.e. increases with distance from the black hole (as discussed elsewhere, this is because the accreting matter ``peels-off" into the wind before approaching the black hole with only a small fraction of it being eventually accreted). So at the outer disk regions ($X \sim 10^5 - 10^6$) the wind mass flux can significantly larger ($\gsim 100$ times) than the accretion rate needed to produce the BH emission, a fact supported by X-ray spectroscopy in a few AGN and Galactic BH. If the dust production is saturated at these large distances, i.e. $\dot M/\dot M_{\rm dust} \simeq 10^{-2}$, we can get the dust mass production rate to exceed that accreted into the black hole.

To investigate further the necessary conditions imposed on the flow parameters (i.e. $M, \dot m, X$) by the formation of dust grains, we compare the timescales for chemical interactions ($t\rm_{chem}$) and the flow-timescale ($t\rm_{flow}$) as a function of the dimensionless distance $X$ from the central source.
The formation of dust in the outflowing wind medium, just like any species that is the product of reactions of the wind constituents, depends on the balance between the flow and the interaction time scales. Therefore one can provide a qualitative view of the parameter space where production is viable and significant.

Considering that the local averaged, chemical dust-producing reaction rate is $\zeta \sim \langle \sigma_{\rm chem} \ v_{\rm th} \rangle$, the chemical evolution time for dust production at a distance $X$ is given by the following expression,

\begin{equation}
\begin{split}
\label{eq_tchem}
t_{\rm chem} & \simeq 1/(n_z (X) \sigma_{\rm chem} v_{\rm th}) \simeq n_z(X)^{-1} \zeta^{-1} \\
& \simeq f_z^{-1} n_0^{-1} \dot m^{-1} X M_8 \ \zeta^{-1} 
\end{split}
\end{equation}
where $n_z$ is the metal density introduced in Section \ref{nucleation} and we have used the density term from Equation \ref{eq_density}. Here we have considered that the reaction rate has a weak dependence on the (dimensionless) distance  $X$ from the BH and ignored the $\theta$ dependance of $n$ for simplicity. For this study, as explain in Section \ref{nucleation}, we have assumed a temperature averaged constant reaction rate $\zeta$ for all nucleation processes.

Similarly, ignoring the $\theta$ dependance, the flow-time $t\rm_{flow}$, using the expression for velocity from Eqn. \ref{eq_vz}, is given by,

\begin{equation}
\begin{split}
\label{eq_tflow_emp}
t_{\rm flow} \sim R/v \sim R/(\epsilon V_K) \sim \epsilon X^{3/2} R_{\rm sch}/c \simeq A X^{3/2}M_8
\end{split}
\end{equation}
where for the expression of velocity we have used Equation \ref{eq_vz}, which shows that the flow-velocity is slower than the local Keplerian velocity. Here we have approximated it by an empirical factor $\epsilon \sim 0.1 \beta \simeq 0.05$, to correct the velocity. The Schwarzschild radius was expressed as 3$\times$10$^{13}$~$M_8$, and the constant A therefore is $\sim 3 \times 10^{13} \epsilon/c$.

Condition for dust formation demands that the timescale for reaction should be shorter than the flow timescale, and hence the ratio should be greater than equal to 1. Using Eqn. \ref{eq_tchem} and \ref{eq_tflow_emp} we can therefore derive the relation, 
\begin{equation}
\label{eq_tflow_treac}
t_{\rm flow}/t_{\rm chem} \ge 1 \ \ \rightarrow \ \ \ C_1 \dot m X^{1/2} \ge 1
\end{equation}
where $C_1$ is the normalization constant appropriate for the time scales and reaction rates introduced in our detailed calculations. Using, $f_z \sim 10^{-4}$, $\zeta \sim 10^{-13}$, $n_0 \sim 5 \times 10^{10}$, we derive that $C_1 \simeq 10^{-4}$.  
Importantly, this signifies that when the gas density normalization is proportional to $\dot m_0 \, M_{\rm BH}^{-1}$ as in our case, the necessary condition for dust formation is independent of the mass of the BH.


Survival of the dust grains at a given $X$ depends on the dust temperature. From the equation for luminosity given by Equation \ref{eq_lum}, we can write, $L \simeq C_2 \dot m M\rm_{BH}$, where $C_2$ is a constant that takes into account the Eddington luminosity and the efficiency $\eta$, ignoring the dependence of luminosity on the inclination angle. For empirical understanding, if we equate this to the blackbody temperature of dust emission, and consider the sublimation temperature for silicate dust as 2000 K, we can derive a limiting condition for the survival of dust grains by,

\begin{equation}
\label{eq_survivalofdust}
C_2 \dot m M_{\rm{BH}} \simeq 4 \pi X^2 R_{\rm{Sch}}^2  \sigma T_{\rm{dust}}^4 \ \rightarrow \ C_2^\prime \dot m X^{-2} M_{\rm{BH}}^{-1} \le 2000^4 \\
\end{equation}
where $\sigma$ is the Stefan - Boltzmann constant. Therefore, the condition necessary for the survival of dust grains does depend on the BH mass $M\rm_{BH}$.


Figure \ref{fig_parameterspace} (right-panel) presents these two constraints ($i.e.$, survival and formation of dust) in terms of $X$ and $\dot m$, directly from Eqns. \ref{eq_tflow_treac} and \ref{eq_survivalofdust}.The parameter space shown in white color on the figure corresponds to the combination of $X$ and $\dot m$ that satisfies both these conditions. It is evident from the figure that with decreasing the mass accretion/outflow rate $\dot m$, dust formation becomes feasible at increasingly larger values of $X$, that is toward the outer boundaries of the accretion disk. Also, the dependence of the black hole mass in these constraints implies that a larger black hole mass makes a larger fraction of $X$-space available to dust formation. As a consequence, a BH of only 10 \Ms, limits the allowed values of $X$ to larger than 10$^7$, (in physical units $R \sim 10^{13}$ cm), which is generally beyond the outer boundary of the accretion disk, suggesting the absence of dust formation in the black holes of the Galactic binaries.

The radial dependence of the winds (taken at present to be proportional to $(\dot m /M) X^{-1}$) affects significantly the rate of dust production in two ways: First by limiting the $\dot m - X$ parameter space where dust formation becomes possible and second by reducing the amount of dust production rate over the volume of the wind. For instance, for $n(X) \propto X^{-3/2}$ \citep{blandford1982}, the ratio of the flow/reaction timescales (the green line of Fig. \ref{fig_parameterspace}b) would be independent of $X$, i.e. it would be a vertical line at some value of $\dot m$. Finally, a radiation pressure driven wind of $n(X) \propto X^{-2}$ would limit even further the parameter space, with the green line having a positive slope, $X^{1/2} \le (\dot m \zeta)$, as one can easily deduce.

The rate of dust production over the wind volume $V$ is proportional to $\dot M_\mathrm{dust}\propto \int n^2(X) \mathrm{d}V(X)$ where the density is expressed as $n(X) \propto (\dot m /M) X^{-p}$. For $p \simeq 1$, $i.e.$, the value taken in our model, we have $\dot M_\mathrm{dust}\propto \dot m^2 M X_{\rm max}$ ($X_{\rm max}$ is the outer boundary of the volume in dimensionless units), a scaling consistent with the result of the detailed calculations depicted in Fig. \ref{fig_parameterspace} (left-panel). For $p=3/2$ \citep{blandford1982}, $\dot M_\mathrm{dust} \propto \dot m^2 M \log X_{\rm max}$ while for $p = 2$, $\dot M_\mathrm{dust} \propto \dot m^2 M (1/ X_{\rm max})$, indicating significant reduction in dust production for the larger values of the parameter $p$, as one intuitively anticipates.


To conclude, we would like to remind the reader that the relations between $\dot m$ and $X$, shown in this section are meant to indicate the  broader trends associated with dust production in AGN winds; they ignore the details of the wind structure and the chemistry involved; however, these broader trends are not in disagreement with the more detailed results shown in fig. \ref{fig_parameterspace} for some specific cases of dust production as a function of $\dot m, M$, for $n(X) \propto X^{-1}$.

\section{Discussion}

\label{sec_conclusion}

This work presents the first attempt to a comprehensive theory of the so-called AGN ``dusty" tori, in that it includes, besides the structure of the tori, in terms of MHD accretion disk winds, also the production of `new dust' in these winds. To date, the presence of any dust has been assumed, constrained only by its absence interior to the dust sublimation line.

The primary objective of the present study has been to determine whether the winds blown off the accretion disks in an AGN can provide an environment that is suitable for new dust formation.

We would like to clarify again that, the term `new dust' is generally associated with the synthesis of new metals in stellar environments through nuclear burning. However, the outflows of AGN accretion disks are not associated directly with any nuclear burning phase. Therefore, in this case, by `new dust' we are referring to the formation of molecules, seeds and dust grains from an initial fully or partly dust-free wind carrying metals synthesized elsewhere, or a high-efficiency of reformation of the dust that has been destroyed by the radiation from the central BH.

The essence of this work is the intimate relation between the nature and structure of the tori themselves, namely the accretion disk MHD winds \citep{konigl1994}, and the formation of the dust they apparently incorporate. To the best of our knowledge this is the first work to employ the details of molecule nucleation and condensation and grain coagulation to study the formation and growth of dust in AGN winds. As discussed in detail, significant amounts of dust can be produced only for winds whose density dependence is sufficiently shallow, i.e. proportional to $X^{-1}$ \citep{conto94,fukumura2010,fukumura2018}.

Assuming the winds have the requisite density profile, one of the main results of our analysis is the quantitative calculation of the (dimensionless) dust mass-flux in these winds, $\dot m_{\rm d}$; this is shown to be  $\dot m_{\rm d}\simeq \dot m^2$ (i.e. the dimensionless mass flux rate onto the black hole), which for $\dot m \simeq 1$, becomes comparable to the mass flux needed to power the black hole emission. This scaling provides a convenient and utile measure of the mass and mass flux of dust in the AGN `dusty tori' in terms of the AGN global properties. As such, it also provides a measure of the rate by which they enrich the AGN interstellar or intergalactic (should they be able to escape the galaxy) medium. Because our models assume no dust present initially in these flows, should observations indicate the presence of more dust than suggested by the above scaling, one would have to surmise that the original flow was not dust-free. The same scaling can also prove useful in following the relative contribution of AGN and star formation to the cosmic dust abundance and therefore to their relative contribution to the cosmic infrared background.

While the winds we have invoked in this work are thought to extend to the black hole vicinity, the inner radius of their dusty segment is much larger, set by the dust sublimation distance (Eq. 24); their outer radius is obtained assuming that the disk terminates at the distance where its Keplerian velocity $V_K(X)$ is a few, say $\lambda$-times, ($\lambda >1$), the dispersion velocity of the overlying spheroid $\sigma$, i.e. $V_K(X_{\mathrm{max}}) \sim \lambda \sigma$. Considering that $V_K / c \simeq X^{-1/2}$, the maximum normalized distance is $X_{\mathrm{max}} \simeq (c/\lambda \sigma)^2$. Assuming further that the BH obeys the $M-\sigma$ relation, i.e. $M \sim 10^7 M_{\odot} (\sigma_2)^4$ or $M_7 \simeq \sigma_2^4$ ($\sigma_2 = \sigma/ 100 \, {\rm km\,s}^{-1})$, we obtain for the case under consideration $R_{\mathrm{max}} \simeq  3 \times 10^{12} M_7 X_{\mathrm{max}} \,{\rm cm} \simeq 10 \, M_7^{1/2} \lambda^{-2}\,{\rm pc}$. This value is consistent with the size of the torus of Cyg A of $\gsim 100$ pc, whose BH mass is $M_8 \gsim 10$, obtained directly by VLA imaging \citep{carilli2019}.

Our calculations have also yielded the dust to gas mass ratio and the size distribution of grains. As shown in the previous section, in the condensation region, the dust to gas mass ratio varies between $10^{-5}$ and $10^{-2}$, increasing with the dimensionless distance $X$. Freshly formed dust is dominated by small size silicate grains, though some grains grow to a few hundred Angstroms. The preponderance of large grains, however, increases with the fraction of initial dust present in the wind.

The presence of grains in these winds should lead to features in the IR AGN spectra. One of the most important such features is the 9.7 $\mu$m silicate feature. This feature should be in emission or absorption, depending on the observer inclination angle, in accordance with the AGN unification scheme \citep{antonucci1993, urry1995}. Such a dependence has been apparently observed in a large sample of a wide range of AGN types (PG QSOs, FR II radio galaxies, Seyfert 1 and 2, LINERS, BAL QSOs etc.) \citep{shi2006, mason2015}, where the inclination angle was estimated by the absorbing X-ray column $N_H$. For small values of $N_H$ ($\log N_H < 21$) the feature appears in emission, while it becomes increasingly prominent in absorption as $\log N_H $ increases beyond $\log N_H > 22 $.

An open issue with the AGN dust emission is that of the presence of clumps. It is generally assumed that the AGN tori are composed of high density clouds similar to those associated with the AGN broad line emission. It is our understanding that this view is sustained by the need of the broad distribution of dust temperatures needed to reproduce the AGN IR spectra: The source covering factor has a radial dependence, which leads to clouds of different temperatures so that they reproduce the observed distribution between 1 and 1000 $\mu$m. It is generally considered that a uniform wind (e.g. \cite{pier1992}) will produce dust emission too hot to be consistent with observations. However, the density profile of our winds allows for a well defined distribution of dust temperatures, given their $1/r$ density distribution, a fact that has been adopted in some models already \citep{rowanrobinson1995}.

Finally, an issue we have not touched upon at all in our model is that of the dynamical effects of radiation pressure on the  dust, considering that its cross section is much larger than that of Thomson.


The radiation pressure force on the dust would exceed that of gravity at luminosities smaller ($\sim 10^{-2}$) than those of Eddington. One could argue that this would modify completely the wind dynamics. While this issue requires much more detailed modeling, we would like to note that the self-similar shape of the poloidal magnetic field lines along which the wind flows, is eventually determined by the pressure of gas in the spheroid into which the BH is located. Its mass is much larger than that of the BH (by a factor of $\sim 10^3$) and it could help the wind maintain its general shape, despite the increased radiation pressure. One, of course, has to bear in mind that our description is confined to the regions of the BH sphere of influence, which is of order of a few parsec. Once driven (by the magnetic fields) beyond this region, the winds are subject to the effects of radiation pressure and would likely expand on all directions in accordance with the influence of the latter. However, it is important that the winds are lift off the accretion disk to heights $H \simeq R_{max}$ before radiation pressure becomes important. This cannot be achieved by the disk radiation pressure near $R_{max}$ whose radiation field momentum is minuscule. The fact that the momentum flux in some of these winds is larger than that of the AGN radiation field, maybe due to the fact that they are indeed driven, at the smaller scales, by the disk magnetic field, as suggested herein.

Other than that, a host of other issues (spectra, variability etc.)  arising with the presence of such detailed models of the dust content of AGN tori, will be addressed in future works.
\section{Summary}
\label{sec_summary}
The results of our paper can be summarized as follows:
\begin{enumerate}
\item We have calculated the dust production rate and the total mass of dust produced in MHD winds driven by a centrally accreting black hole (BH) characterized by a $M_{\rm BH}$, and a dimensionless accretion rate $\dot{ m}$. 
\item We assumed that the accreted gas is dust free, and solar composition, and that the AGN luminosity is the only radiative source controlling the temperatures of the gas and dust in the outflow. The formation of dust starts with chemical reactions that form seed nuclei which grow by coagulation and accretion throughout the flow. Since the inflowing gas is oxygen rich (C/O $<$ 1), the seed nuclei consist of various magnesium and silicate oxides, which grow by accretion and coagulation to silicate grains.  
\item We identify four regions around the AGN, shown in Figure \ref{fig_zones}: (a) sublimation  zone where the temperature of any seed nuclei exceeds the sublimation temperature; (b) a region where seeds can form but do not accrete or coagulate to form dust grains; (c) a region where seeds can form and grow; and (d) a region where no new seeds can form, but existing seeds can grow by accretion and coagulation. 
\item We show that the resultant distribution of newly formed dust resembles a toroidal volume of dusty gas, which we then identify as the dusty torus. Morever the size of the torus matches well with that estimated from VLA imaging. We emphasize that the dusty torus is therefore not a static region of dusty gas surrounding the AGN, but instead a dynamic phenomena produced by the outflowing winds. 
\item Formation of new dust, characterized by the high column density and opacities along equatorial lines of sight relative to the polar ones, are in agreement with the role of dust associated with the AGN unification theory. 
\item We justify the importance of $r^{-1}$ density law in the winds, for the efficient formation and survival of dust grains, supporting the MHD models where this density law was adopted. 
\item We find that the dust production rate can be approximated by a law such as $\dot M_\mathrm{dust} \simeq A \ \dot m^{2.25} M_8$, where is $A$ constant given by $\sim$ 3.5~yr$^{-1}$. This is an upper limit. 
\item The production rates and dust yields presented above assume that the material accreted by the BH is initially dust free. If a fraction of the condensable material, $f_d$, was initially locked up in dust then the {\it net} dust production rates need to be multiplied by $1-f_d$. In that case, the AGN may be an important processor of any pre-existing dust. Their sublimation and reformation will alter their composition and size distribution. 
\item The relative contribution of AGNs and star forming galaxies to the cosmic abundance of dust can be related to their relative contribution to the cosmic infrared background, and will require further detailed studies.
\item The dust blown out of the galaxy by the AGN may be injected to the intergalactic medium, and contribute to the diffuse IR emission observed in galaxies.
\end{enumerate}

We acknowledge NASA’s 16-ATP2016-0004 grant for supporting this project. We thank the anonymous referee for the valuable suggestions that have helped us in finalizing the manuscript of this paper.

\software{CLOUDY \citep{fer98}}

\bibliographystyle{aasjournal.bst}
\bibliography{Bibliography_sarangi}

\end{document}